\documentclass[12pt]{article}

\usepackage{amsmath, amsfonts, amssymb, amsthm, bm}
\usepackage{mathtools}
\usepackage{bbm}
\usepackage{natbib}
\usepackage{url} 
\usepackage{dirtytalk}
\usepackage{comment}
\usepackage{hyperref}
\hypersetup{
  pdftitle={Title},
  pdfauthor={Author 1; Author 2},
  pdfkeywords={3 to 6 keywords, that do not appear in the title},
  colorlinks=true,
  linkcolor={blue},
  filecolor={Maroon},
  citecolor={Blue},
  urlcolor={Blue},
  pdfcreator={LaTeX via pandoc}}

\usepackage[utf8]{inputenc}

\newtheorem{prop}{Proposition}
\newtheorem{cor}{Corollary}
\theoremstyle{definition}
\newtheorem{definition}{Definition}

\newtheorem{corollary}{Corollary}

\usepackage[dvipsnames]{xcolor}
\usepackage{graphicx}
\usepackage{subcaption}

\usepackage{enumitem}

\newcommand{\Beta}{\text{Beta}}
\newcommand{\N}{\text{N}}
\newcommand{\E}{\mathbb{E}}

\newcommand{\Bern}{\text{Bern}}

\newcommand{\simiid}{\overset{\mathrm{iid}}{\sim}}
\newcommand{\simind}{\overset{\mathrm{ind}}{\sim}}
\renewcommand{\P}{\mathrm{Pr}}

\newcommand{\blind}{0}

\allowdisplaybreaks

\usepackage{xr}
\begin{document}

\def\spacingset#1{\renewcommand{\baselinestretch}%
{#1}\small\normalsize} \spacingset{1}

\date{}

\if0\blind
{
  \title{\bf Dependent Dirichlet processes via thinning}
  \author{Laura D'Angelo, Bernardo Nipoti, and Andrea Ongaro 
  \hspace{.2cm}\\
 University of Milano-Bicocca, Italy}
  \maketitle
} \fi

\if1\blind
{
  \begin{center}
  \vspace*{1.6cm}
  {\LARGE\bf Dependent Dirichlet processes via thinning}
\end{center}
  \bigskip
} \fi

\bigskip
\begin{abstract}
We introduce a novel framework to model a collection of samples using dependent Dirichlet processes constructed through a thinning mechanism. The proposed approach modifies the stick-breaking representation of the Dirichlet process by thinning, that is, setting equal to zero a random subset of the beta random variables used in the original construction. This results in a collection of dependent random distributions that exhibit both shared and unique atoms, with the shared ones assigned distinct weights in each distribution. The generality of the construction allows expressing a wide variety of dependence structures among the elements of the generated random vectors. Moreover, its simplicity facilitates the characterization of several theoretical properties and the derivation of efficient computational methods for posterior inference. A simulation study illustrates how a modeling approach based on the proposed process reduces uncertainty in group-specific inferences while preventing excessive borrowing of information when the data indicate it is unnecessary. This added flexibility improves the accuracy of posterior inference, outperforming related state-of-the-art models. An application to the Collaborative Perinatal Project data highlights the model's capability to estimate group-specific densities and uncover a meaningful partition of the observations, both within and across samples, providing valuable insights into the underlying data structure.
\end{abstract}

\noindent%
\vfill

\newpage
\spacingset{1.2} 

\section{Introduction}

Over the past 25 years, considerable attention has been dedicated to the study of dependent nonparametric priors for collections of random probability measures indexed by a covariate. A recent review by \citet{Quintana2022} highlights many key contributions in this area. \citet{Fra25} introduce a unifying framework that clarifies how these contributions fit within the broader class of multivariate species sampling models. A general distribution theory of hierarchical processes is presented in \citet{Cam19}. The dependent Dirichlet process (DDP), introduced by \citet{maceachern2000, maceachern2026} in a seminal paper, remains arguably the most popular prior of this type. The DDP extends the use of the Dirichlet process (DP) to nonparametric regression, enabling the modeling of the conditional distribution of the response given some covariate. The definition of the DDP builds on the stick-breaking representation of the DP, as introduced in \citet{sethuraman1994}. This constructive definition characterizes a random distribution $p$ on a measurable space $(\Theta,\mathcal{B})$ starting from two sequences of independent and identically distributed (iid) random variables $\{v_j\}_{j\geq1}$ and $\{\theta_j\}_{j\geq 1}$, independent of each other. Assuming $v_j \simiid \Beta(1,\alpha)$ with $\alpha >0$, and $\theta_j\simiid P_0$, where $P_0$ is a probability distribution on $(\Theta,\mathcal{B})$, for $j\geq 1$, \citet{sethuraman1994} showed that a DP with base measure $P_0$ and total mass $\alpha$ can be expressed as
\begin{equation}\label{eq:DP}
	p(\cdot) = \sum_{j=1}^{\infty} \omega_j \delta_{\theta_j}(\cdot),
\end{equation}
where $\delta_{\theta_j}$ indicates a Dirac mass at the location $\theta_j$, and the probabilities are defined as $\omega_1 = v_1$, $\omega_j = v_j \prod_{h<j}(1-v_h)$ for $j>1$. The DDP extends this formulation by defining a collection of random probability measures $p_x$, indexed by a covariate 
$x\in\mathcal{X}$, through modifications to the stick-breaking representation of the DP. Specifically, either the weights, the atoms, or both are replaced with stochastic processes indexed by $x$. This construction preserves the desirable property that the marginal distribution at each covariate value remains a DP. Moreover, when the covariate indexing the collection of random probability measures is categorical, say $g$ with levels $\{1, \dots, G\}$, case considered throughout this work, the DDP serves as a convenient prior for modeling data structured into distinct yet related groups, e.g., patients undergoing different treatments. A notable instance of MacEachern’s DDP is the single-atom DDP. As the name suggests, this specification maintains a common set of atoms across all covariate values, introducing dependence solely through the weights. This formulation has been widely studied, particularly for modeling dependence over time \citep{griffin2011} and across spatial locations \citep{duan2007, fuentes2013}.  A related approach was proposed by \citet{griffin2006}, with the dependence between DPs introduced by modifying the ordering of the beta-distributed random variables in their stick-breaking construction. 
Models for dependent probability measures based on the idea of sharing some or all the atoms extend beyond the DDP framework. The hierarchical Dirichlet process \citep[HDP,][]{Teh2006} models a collection of probability distributions as distinct DPs, introducing dependence and shared atoms by modeling the common base measure as a DP itself. The common atoms model \citep[CAM,][]{Denti2023} employs a nested DP structure with a shared set of atoms to jointly cluster groups defined by covariate values and individual observations. A different strategy for modeling dependence is proposed by \cite{muller2004}, who combine common and idiosyncratic random probability measures to capture shared and sample-specific behaviors. This distinction between common and idiosyncratic components is also central to the Griffiths--Milne dependent Dirichlet process \citep[GM-DDP,][]{GMDDP}, which exploits the normalization of dependent completely random measures (CRMs) to define a vector of dependent DPs as the superposition of component-specific processes and a shared one. Finally, \cite{compound} obtain a vector of dependent random probability measures by normalizing compound CRMs sharing a common sequence of locations.

We introduce the thinned dependent Dirichlet process (thinned-DDP), a novel nonparametric prior for modeling collections of dependent random probability measures $(p_1,\ldots,p_G)$. This prior is derived by thinning the atoms of a common DP. Instead of directly selecting the atoms $\theta_j$ that contribute to the definition of a covariate-specific random probability measure $p_g$, the approach we propose modifies the weights $\omega_j$ in \eqref{eq:DP} by multiplying the common random variables $v_j$  by covariate-specific thinning variables $\ell_{j,g}\in\{0,1\}$, which determine whether the corresponding atoms $\theta_j$ appear in the definition of $p_g$. This approach retains the simplicity and tractability of the original stick-breaking construction, enabling analytical study of the properties of the thinned-DDP, and facilitating the adaptation of computational methods originally developed for a single DP to multi-sample settings. 
Unlike the single-atom DDP, the HDP, and the CAM, where all atoms are shared by all distributions, the thinned-DDP allows atoms to be either shared among subsets of measures or specific to individual ones. 
Moreover, the thinned-DDP offers a more flexible dependence structure compared to the GM-DDP. While both approaches account for the presence of shared and non-shared atoms, the thinned-DDP, unlike the GM-DDP, allows atoms to be shared by specific subsets of component-specific processes rather than by all of them. Finally, we note that the idea of modifying a nonparametric process through thinning, that is, by skipping some of its atoms, has been used in some recent contributions: \citet{lau_cripps} introduced thinned CRMs to model dependent hazard functions; the quasi-Bernoulli stick-breaking process of \citet{zeng2023} uses a related idea to improve estimation of the number of clusters; and \citet{bi2023} applied a thinning mechanism to the HDP to explicitly identify population-specific and shared atoms.

The remainder of the paper is organized as follows. Section~\ref{sec::definition} introduces the thinned-DDP and studies its prior properties. Section~\ref{sec:ESA} presents a notable instance, the eventually single-atom thinned-DDP. Section~\ref{sec:prior} discusses two prior specifications for the thinning variables and their effects on the process. Section~\ref{sec::posterior_inference} defines a mixture model with thinned-DDP-distributed mixing measures and outlines a blocked Gibbs sampler for posterior inference. Section~\ref{sec::simulation} reports a simulation study, including comparisons with state-of-the-art models. Section~\ref{sec::application} applies the proposed model to gestational-age data from women in different hospitals. Proofs and additional results are provided in the Supplementary Material.


\section{Dependent Dirichlet processes via thinning}\label{sec::definition}
We let $g$ be a categorical covariate with a finite number of levels, say $\{1,\ldots,G\}$. In addition to the components underlying the stick-breaking construction of a DP given in~\eqref{eq:DP}, namely $v_j \simiid \text{Beta}(1, \alpha)$ with $\alpha > 0$ and $\theta_j \simiid P_0$ for $j \geq 1$, 
we introduce $G$ sequences of thinning variables $\{ \ell_{j,g} \}_{j \geq 1}$, for $g = 1, \dots, G$, where $\ell_{j,g}\in\{0,1\}$ and $\sum_{j=1}^\infty \ell_{j,g} = \infty$. We let $\ell_{1:G}$ denote the collection of these sequences. For each realization of the covariate $g\in\{1,\ldots,G\}$, we define a random probability measure $p_g$ on $(\Theta,\mathcal{B})$ as
\begin{equation}
	p_g(\cdot) = \sum_{j=1}^{\infty} \omega_{j,g}\delta_{\theta_j}(\cdot) = \sum_{j=1}^{\infty} \bigg[v_j\ell_{j,g} \prod_{h<j}(1-v_h\ell_{h,g})\bigg] \delta_{\theta_j}(\cdot).
 \label{eq::thinnedDDP}
\end{equation}
\begin{definition}\label{def:thinned-DDP}
    A thinned dependent Dirichlet process with concentration parameter $\alpha>0$, base probability measure $P_0$, and thinning sequences $\ell_{1:G}$, is a vector of random probability measures $p_{1:G}=(p_1,\ldots,p_G)$ where each $p_g$, with $g=1,\ldots,G$, is defined as in~\eqref{eq::thinnedDDP}. We introduce the notation $p_{1:G}\sim \text{thinned-DDP}(\alpha,P_0,\ell_{1:G})$. 
\end{definition}
Definition~\ref{def:thinned-DDP} introduces a hierarchical construction for the vector $p_{1:G}$, where each component is a modified version of a shared DP $p$, introduced in~\eqref{eq:DP}, henceforth called the parent process. This framework offers a hierarchical alternative to the well-known HDP \citep{Teh2006}, with a crucial difference: in the thinned-DDP, each marginal $p_g$ is itself a DP with known parameters $\alpha$ and $P_0$. In contrast, in the HDP, each element of the analog of $p_{1:G}$, while distributed as a DP conditionally on the realization of a common DP-distributed process, is characterized by a base measure that depends on that realization. The covariate-specific sequences $\ell_{1:G}$ act as a thinning mechanism that modifies the stick-breaking construction of the parent process. Specifically, when $\ell_{j,g} = 0$, the corresponding weight $\omega_{j,g}$ is set to zero, and the atom $\theta_j$ is excluded from the construction of $p_g$. Using the stick-breaking metaphor from the definition of a DP, we can interpret $\omega_{j,g}=0$ as indicating that, at the $j$th step, the stick used to define $p_g$ is not broken. The resulting unassigned probability mass is redistributed by scaling all subsequent weights in $p_g$ by a factor of $(1 - v_j)^{-1}$, ensuring a well-defined random probability measure. The elements of $p_{1:G}$ are thus defined through the same set of atoms  $\{\theta_j\}_{j \geq 1}$, but assign them different probabilities. This aligns with the single-atom DDP framework of \citet{maceachern2000}, with the key distinction that, in the thinned-DDP construction, certain weights are exactly zero, a mechanism that explicitly models whether atoms are covariate-specific or shared. 
The thinned-DDP can thus be regarded as a limiting case of the single-atom DDP. Accordingly, conditionally on the $\ell_{1:G}$, the random probability measures $p_{1:G}$ are identically distributed as a DP with concentration parameter $\alpha$ and base measure $P_0$, the marginal distribution of each $p_g$ being invariant to $\ell_g$. The condition $\sum_{j=1}^\infty \ell_{j,g} = \infty$ indeed ensures that the unit stick is broken infinitely many times, with the weight generated the $j$th time the stick is actually broken 
having the same distribution as $\omega_j$ in~\eqref{eq:DP}. Although the focus of the present paper is on the case of a finite collection of groups, a fully continuous-index version of the thinned-DDP, closer in spirit to MacEachern's DDP, could be obtained by replacing the group-specific thinning sequences $\{\ell_{j,g}\}_{j\geq 1}$ with covariate-indexed binary processes $\{\ell_j(x):x\in\mathcal{X}\}_{j\geq 1}$. In this way, random probability measures would be constructed directly as functions of the continuous covariate $x$, with dependence across covariate values induced by the joint distribution of the processes $\ell_j(\cdot)$.
Finally, we observe that the thinned-DDP can also be framed as an instance of the general class of multivariate species sampling processes introduced by \citet{Fra25}.

We examine the mixed moments of a thinned-DDP, given the sequences $\ell_{1:G}$, to explore the dependence between pairs of elements in $p_{1:G}$. 
Proposition~\ref{prop:corr} characterizes the correlation between two random distributions $p_1$ and $p_2$ jointly distributed as a thinned-DDP. 
$\mathbb{I}_{\{A\}}$ denotes the indicator function, which equals 1 if the condition $A$ holds, and 0 otherwise.
\begin{prop}\label{prop:corr}
If $p_{1:2}\sim\text{thinned-DDP}(\alpha,P_0,\ell_{1:2})$, then, for any $A\in\mathcal{B}$, the correlation between $p_{1}(A)$ and $p_{2}(A)$ is given by 
\begin{equation*}
    \mathrm{Corr}\left(p_1(A), p_2(A) ; \ell_{1:2} \right)=\frac{2}{(\alpha+2)} \sum_{j\geq 1}  \ell_{j,1}\ell_{j,2} \left(\frac{\alpha}{\alpha+2} \right)^{s_j} \left(\frac{\alpha}{\alpha+1} \right)^{q_j} ,
\end{equation*}
where $s_j = \sum_{h=1}^{j-1}\ell_{h,1}\ell_{h,2}$ and $q_j = \sum_{h=1}^{j-1}\mathbb{I}_{\{\ell_{h,1}\neq \ell_{h,2}\}}$ are, respectively, the number of shared and distribution-specific atoms in the first $j-1$ components of $\ell_{1:2}$. 
\end{prop}
The correlation in Proposition~\ref{prop:corr} is expressed as an infinite sum, reflecting its dependence on the infinite sequences $\ell_{1:2}$. Within this sum, only the terms where both $\ell_{j,1}$ and $\ell_{j,2}$ are equal to one contribute to this quantity. 
Such correlation is always non-negative and spans the entire interval $[0,1]$, attaining one when $\ell_1$ and $\ell_2$ are identical, and zero when no position contains a one in both sequences. Additionally, the expression in Proposition~\ref{prop:corr} is independent of the choice of $A$, a measurable subset of $\Theta$, which motivates the use of this correlation as a measure of dependence between the random probability measures $p_1$ and $p_2$. This aligns with the findings of \citet{Fra25}, who showed that the same holds for the whole class of multivariate species sampling models. The result in Proposition~\ref{prop:corr} is readily adapted to study the correlation between a generic component of $p_{1:2}$ and the parent process $p$; see Corollary~\ref{cor:corr_parent} in the Supplementary Material.

We conclude this section by analyzing the prior number of distinct values, or clusters, implied by a thinned-DDP model. The thinned-DDP allows for both common and covariate-specific atoms $\theta_j$. Consequently, samples drawn from a thinned-DDP may share distinct values while also exhibiting values unique to specific components. To investigate this, we consider two samples, $y_{1,1},\ldots,y_{n_1,1} \mid p_{1:2} \simiid p_1$ and $y_{1,2},\ldots,y_{n_2,2} \mid p_{1:2} \simiid p_2$, where $p_{1:2} \sim \text{thinned-DDP}(\alpha,P_0,\ell_{1:2})$ for some pair of sequences $\ell_{1:2}$. We define $K_0$ as the number of distinct values shared between the two samples and let $K_1$ and $K_2$ represent the number of distinct values unique to the first and second samples, respectively. The total number of distinct values across both samples is thus given by $K = K_0 + K_1 + K_2$. To emphasize the dependence of $K$ on $n_1$, $n_2$, and $\ell_{1:2}$, we introduce the notation $K(n_1,n_2;\ell_{1:2})$, which we use when convenient. We next focus on the prior expected value of $K$: while its exact form depends on the sequences $\ell_{1:2}$, it is nonetheless possible to establish bounds for it. The next result formalizes the intuitive idea that the expected total number of distinct values in two samples from a thinned-DDP falls between two extremes: (i) the expected number of distinct values in a single DP-distributed sample whose size equals the combined sizes of the two samples, and (ii) the sum of the expected number of distinct values in two independent DP-distributed samples.
\begin{prop}\label{prop:exact_K}
Let $y_{1,1},\ldots,y_{n_1,1} \mid p_{1:2} \simiid p_1$ and $y_{1,2},\ldots,y_{n_2,2} \mid p_{1:2} \simiid p_2$, where $p_{1:2} \sim \text{thinned-DDP}(\alpha,P_0,\ell_{1:2})$. Then
\begin{equation*}
        \sum_{i=1}^{n_1+n_2}\frac{\alpha}{\alpha+i-1}\leq \mathbb{E}\left[K(n_1,n_2;\ell_{1:2})\right]\leq\sum_{i=1}^{n_1}\frac{\alpha}{\alpha+i-1}+\sum_{i=1}^{n_2}\frac{\alpha}{\alpha+i-1}.
    \end{equation*}
\end{prop}
By combining  Proposition~\ref{prop:exact_K} with the asymptotic analysis of the number of distinct values in samples from a DP \citep{Kor73}, we find that the growth of $\mathbb{E}[K(n_1, n_2; \ell_{1:2})]$, as $n_1 = n_2 = n \to \infty$, is bounded below by $\alpha\log(n)$ and above by $2\alpha\log(n)$, with the bounds being sharp.

\subsection{Eventually single-atom thinned-DDP}\label{sec:ESA}
The thinned-DDP class has been introduced in its full generality, defined conditionally on $G$ infinite sequences of thinning variables. However, simpler specifications can be achieved by imposing constraints on the structure of these sequences. For example, one may consider thinning sequences that eventually become constantly equal to one. Within this framework, a particularly simple case arises when each sequence $\ell_g$ consists of $u_g - 1$ initial zeros followed by an infinite sequence of ones.
As a result, each $p_g$ is defined as
\begin{equation*}
p_g(\cdot) = \sum_{j=u_g}^\infty \left[v_{j} \prod_{h=u_g}^{j-1}(1-v_{h}) \right] \delta_{\theta_j}(\cdot),
\end{equation*}
where $u_g$ is, for any $g=1,\ldots,G$, a positive integer. We refer to the resulting distribution of the vector $p_{1:G}$ as the eventually single-atom thinned-DDP, reflecting the observation that all atoms $\theta_j$ with $j \geq \max(u_1, \ldots, u_G)$ are shared among all components of $p_{1:G}$, and we introduce the notation $p_{1:G}\sim\text{thinned-DDP}(\alpha,P_0,u_{1:G})$, where $u_{1:G}=(u_1,\ldots,u_G)$. This formulation implies a nested structure among the distributions. For example, if $G = 3$ and $u_1 < u_2 < u_3$, then $p_3$ includes only common atoms, $p_2$ includes an additional $u_3 - u_2$ atoms, shared with $p_1$ but not in $p_3$, and $p_1$ includes a further $u_2 - u_1$ atoms unique to itself. The intuition behind this definition is that differences among distributions can be flexibly captured by the initial atoms in $\{\theta_j\}_{j \geq 1}$, which are those assigned the highest expected probabilities.  The simplicity of the eventually single-atom thinned-DDP is evident in the remarkably simple form that the correlation in Proposition~\ref{prop:corr} takes in this case. This is illustrated in the following result.
\begin{cor}\label{corollary:corr_eventually}
If $p_{1:2}\sim\text{thinned-DDP}(\alpha,P_0,u_{1:2})$, then, for any $A\in\mathcal{B}$, the correlation between $p_{1}(A)$ and $p_{2}(A)$ is given by
\begin{equation*}
	\mathrm{Corr}(p_1(A),p_2(A); u_{1:2}) = 
    \left(\frac{\alpha}{\alpha+1}\right)^{\lvert u_2-u_1 \rvert}.
\end{equation*}
\end{cor}
The correlation depends on the sequences of thinning variables solely through the absolute difference between $u_1$ and $u_2$, which represents the number of atoms not shared by $p_1$ and $p_2$. Moreover, when $u_1 = u_2$, the correlation equals one, as $p_1 = p_2$. Conversely, as $|u_2 - u_1| \to \infty$, the correlation approaches zero, since the shared atoms contribute negligible probability in defining one of the distributions. Under the same assumptions, we can study the correlation between a component of $p_{1:2}$ and the parent process $p$; see Corollary~\ref{cor:parent_eventually} in the Supplementary Material.\\
Finally, the eventually single-atom thinned-DDP structure also allows us to derive an explicit expression for the prior expectation of the number of distinct values $K$, already studied in Proposition~\ref{prop:exact_K} for a generic thinned-DDP.  This is formalized in the next proposition: although the expression may seem complex, its numerical evaluation is straightforward.
\begin{prop}\label{prop:exact_K_eventually}
Let $y_{1,1},\ldots,y_{n_1,1} \mid p_{1:2}\simiid p_1$ and $y_{1,2},\ldots,y_{n_2,2} \mid p_{1:2} \simiid p_2$, where $p_{1:2} \sim \text{thinned-DDP}(\alpha,P_0,u_{1:2})$. If we assume that $u_1\leq u_2$, then
    \begin{align*}
        \mathbb{E}[K(n_1,n_2,u_{1:2})]&=\sum_{r=1}^{n_1}(-1)^{r-1}\binom{n_1}{r} \frac{\Gamma(\alpha+1)\Gamma(r)}{\Gamma(\alpha+r)}\left(1-\left(\frac{\alpha}{\alpha+r}\right)^{u_2-u_1}\right)\\
        &+\sum_{r=0}^{n_1}\binom{n_1}{r}\left(\sum_{i=1}^{n_2+n_1-r}\frac{\alpha}{\alpha+i-1}\right)\sum_{l=0}^r (-1)^l \binom{r}{l} \left(\frac{\alpha}{\alpha+l+n_1-r}\right)^{u_2-u_1}.
    \end{align*}
\end{prop}

We emphasize that although this section presents a specific example, the eventually single-atom thinned-DDP admits more general tractable formulations. One such formulation is described in Section~\ref{subsec:symmetric_ev} of the Supplementary Material.

\section{Priors for the thinning sequences}\label{sec:prior}
For practical application as a Bayesian model with $G$ samples, it is convenient to combine the thinned-DDP with a prior distribution for the sequences $\ell_{1:G}$. Such a prior should be specified to ensure the flexibility of the resulting model. A particularly desirable property of a nonparametric model is its large support, which ensures that the model does not concentrate its probability mass on small subsets of probability measures. We next demonstrate that, under mild conditions on the prior distribution for $\ell_{1:G}$, the resulting thinned-DDP has full weak support. Conveniently, this implies that introducing dependence among the components of a vector of nonparametric priors via a thinned-DDP distribution preserves the flexibility of the DP in estimating joint and marginal distributions. \textcolor{black}{Formally, we let $\mathcal{P}(\Theta)$ denote the set of all probability measures defined on the Borel measurable space $(\Theta, \mathcal{B})$, with $\Theta \subseteq \mathbb{R}^q$ and $q \in \mathbb{N}$. We further assume that the base measure $P_0$ has full support $\Theta$ and let $\mathcal{P}(\Theta)^G$ be the space of all $G$-dimensional vectors whose components take values in $\mathcal{P}(\Theta)$. We define $\mathcal{B}(\mathcal{P}(\Theta)^G)$ as the Borel $\sigma$-field generated by the product topology of weak convergence. The weak support of the thinned-DDP $p_{1:G}$ is then the smallest closed set in $\mathcal{B}(\mathcal{P}(\Theta)^G)$ to which $p_{1:G}$ assigns probability one.} The next proposition establishes that, under mild conditions on the prior distribution on the thinning sequences, the weak support of the thinned-DDP coincides with $\mathcal{P}(\Theta)^G$. This result extends the comprehensive study of the support of the DDP by \citet{Barrientos2012} to the thinned-DDP case, which, as mentioned, can be seen as a limiting case of the class of DDP. Although the result is presented for the case $G=2$, the extension to larger values of $G$ is straightforward.
\vspace{-.1cm}
\begin{prop}\label{prop::support}
Let $p_{1:2}\sim\text{thinned-DDP}(\alpha,P_0,\ell_{1:2})$ \textcolor{black}{be such that $P_0$ has full support  $\Theta$}, and assume that the sequences $\ell_{1:2}$ are assigned a prior distribution, independent of the sequences $\{\theta_j\}_{j\geq 1}$ and $\{v_j\}_{j\geq 1}$, that, for every $k\in\mathbb{N}$, assigns positive probability to the set
\begin{equation}\label{eq::omega1}
 \Omega_k = \{ \ell_{j,1} = 1,
\ell_{j,2} = 0 \: \text{for $j=1,\dots,k$ and }  \ell_{j,2} = 1 \: \text{for $j=k+1,\dots,2k$} \}.
\end{equation}
Then $\mathcal{P}(\Theta)^G$ is the weak support of the process, i.e., the thinned-DDP has full weak support.
\end{prop}

Condition~\eqref{eq::omega1} is very general and typically easily verifiable, allowing for various prior specifications that satisfy it. In Sections~\ref{sec:bern} and~\ref{sec:Poisson}, we thoroughly discuss two specific prior distributions for $\ell_{1:G}$ that meet this condition and thus lead to Bayesian models with full weak support. 
Two additional prior distributions are introduced in Sections~\ref{subsec:dependent_bernoulli} and~\ref{subsec:blocked_poisson} of the Supplementary Material.

\subsection{Bernoulli thinning}\label{sec:bern}
A straightforward specification for the prior distribution of $\ell_{1:G}$ assumes independent Bernoulli priors for the elements of each covariate-specific sequence. Specifically, we let $\ell_{j,g} \simiid \Bern(\pi_g)$, with $\pi_g \in (0,1)$, for $j \geq 1$ and $g = 1, \dots, G$.
We next characterize the correlation between pairs of distributions, marginalizing the thinning sequences. 
\begin{prop}\label{prop:corr_bernoulli}
 Let $p_{1:2}\sim\text{thinned-DDP}(\alpha,P_0,\ell_{1:2})$ and consider a measurable set $A\in\mathcal{B}$. Assume $\ell_{j,1}\simiid \Bern(\pi_1)$ and $\ell_{j,2}\simiid \Bern(\pi_2)$ independent of each other, for $j \geq 1$. The correlation between $p_1(A)$ and $p_2(A)$ is
\begin{equation*}
	\mathrm{Corr}(p_1(A),p_2(A))  =
    \frac{2\, \pi_1\pi_2 (\alpha+1) }{\alpha(\pi_1+\pi_2)+2(\pi_1 + \pi_2-\pi_1\pi_2)} .
\end{equation*}
Moreover, in the case $\pi_1 = \pi_2 = \pi$, the correlation simplifies to 
$$\mathrm{Corr}(p_1(A),p_2(A)) = \frac{\pi (\alpha+1)}{\alpha + 2 - \pi}.$$
\end{prop}
We observe that when $\pi_1=\pi_2=1$, the two distributions are equal, and the correlation is one. \textcolor{black}{As $\pi_1=\pi_2=\pi$ gets smaller, the probability of discarding atoms increases, and so does the probability of creating non-shared components, leading to a weaker correlation.} \textcolor{black}{Corollary~\ref{cor:corr_parent_bernoulli} in the Supplementary Material reports the correlation between a component of $p_{1:2}$ and the parent process $p$, under the same prior specification for $\ell_{1:2}$.}

\begin{figure}[t!]
  \centering
  \includegraphics[width=.9\linewidth]{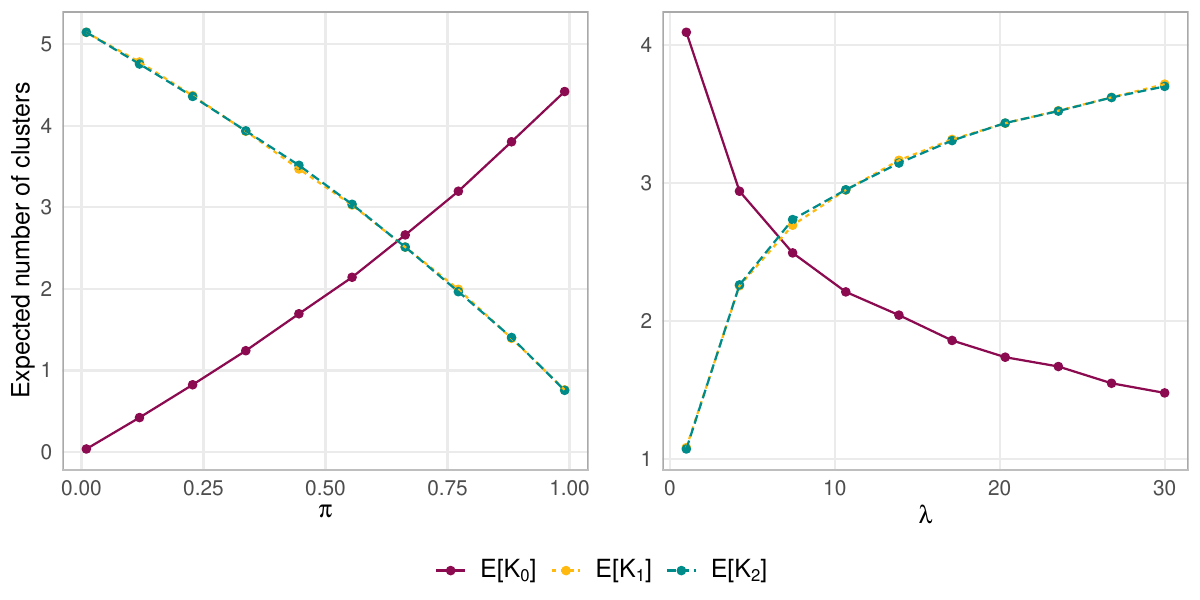}
  \caption{Expected number of shared clusters ($\mathbb{E}[K_0]$, continuous line) and covariate-specific clusters ($\mathbb{E}[K_1]$ and $\mathbb{E}[K_2]$, dashed lines) in two samples of size $n=100$ from $p_1$ and $p_2$, with $p_{1:2}\sim\text{thinned-DDP}(1,P_0,\ell_{1:2})$. Left: Bernoulli thinning, with $\pi$ ranging in $(0,1)$; Right: eventually single-atom Poisson thinning, with $\lambda$ ranging in $(0,30)$.}
  \label{fig:expected_number_atoms}
\end{figure}

\begin{figure}[h!]
  \centering
  \includegraphics[width=0.45\linewidth]{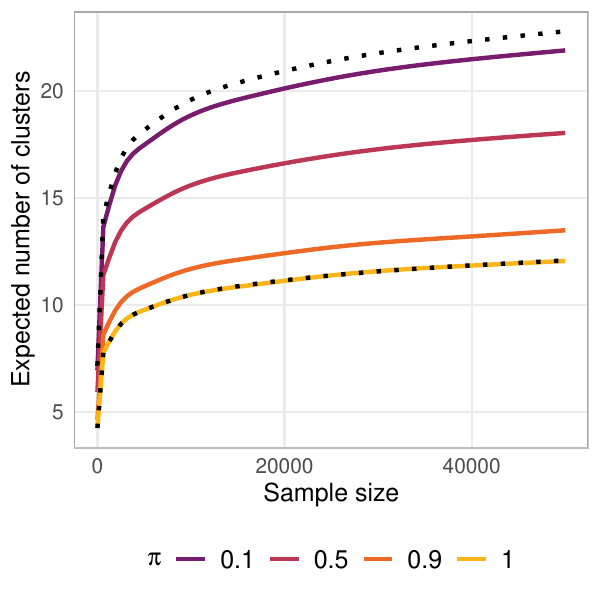}
  \hspace{0.05cm}\includegraphics[width=0.45\linewidth]{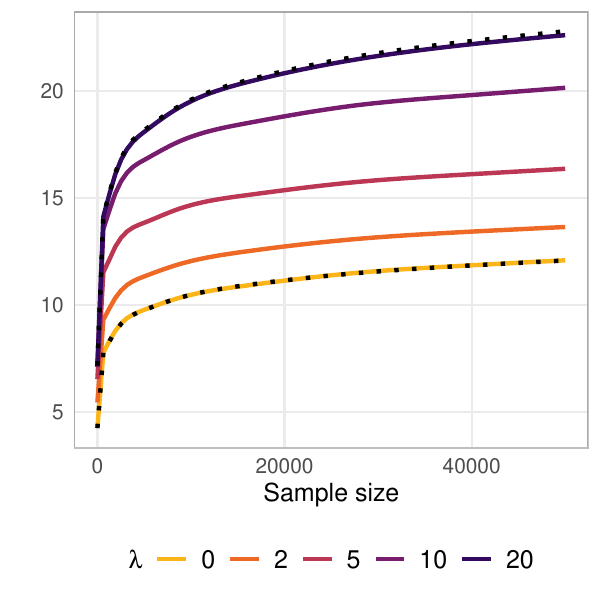}
  \caption{Total expected number of clusters, $\mathbb{E}[K]$, in two samples from $p_1$ and $p_2$, with $p_{1:2}\sim\text{thinned-DDP}(1,P_0,\ell_{1:2})$, as a function of the size $n$ of the two samples. Left: Bernoulli thinning, with $\pi$ ranging in $\{0.1,0.5,0.9,1\}$; Right: eventually single-atom Poisson thinning, with $\lambda$ ranging in $\{0,2,5,10,20\}$. The dotted lines correspond to the theoretical bounds provided by Proposition~\ref{prop:exact_K}.}
  \label{fig:tot_expected_number_atoms}
\end{figure}

We now present a numerical investigation to assess how assigning a Bernoulli prior to the thinning sequences in a thinned-DDP affects the prior distribution of the number of distinct values, both shared across components and component-specific. Our analysis focuses on the case of $G = 2$ samples of size $n$ from a vector $p_{1:2} \sim \text{thinned-DDP}(1, P_0, \ell_{1:2})$, with $\ell_{j,g} \simiid \text{Bern}(\pi)$ for $g = 1,2$. The left panel of Figure~\ref{fig:expected_number_atoms} shows Monte Carlo estimates of $\mathbb{E}[K_0]$, $\mathbb{E}[K_1]$, and $\mathbb{E}[K_2]$ when $n=100$ and for various values of the common thinning probability $\pi$ in $(0,1)$. When $\pi$ is small, the expected number of shared clusters $\mathbb{E}[K_0]$ is close to zero. This occurs because both $p_1$ and $p_2$ are constructed by skipping most of the $\theta_j$ in~\eqref{eq::thinnedDDP}, resulting in few atoms shared by $p_1$ and $p_2$. As $\pi$ increases, more atoms are common to $p_1$ and $p_2$, leading to larger values of $\mathbb{E}[K_0]$. In contrast, the covariate-specific expected number of clusters, $\mathbb{E}[K_g]$ for $g = 1,2$, exhibits the opposite behavior: since each $p_g$ is marginally distributed as a DP with concentration parameter $\alpha = 1$, the estimated total expected number of distinct clusters within each sample, i.e. $\mathbb{E}[K_0] + \mathbb{E}[K_g]$, is approximately equal to $\sum_{i=1}^{100} 1/i \approx 5.19$ for all $\pi \in (0,1)$. 
The left panel of Figure~\ref{fig:tot_expected_number_atoms} illustrates the growth of the total expected number of clusters, $\mathbb{E}[K] = \mathbb{E}[K_0] + \mathbb{E}[K_1] + \mathbb{E}[K_2]$, as the sample size $n$ increases. 
Interestingly, as $\pi$ approaches the extremes of the interval $(0,1)$, $\mathbb{E}[K]$ gets close to the bounds established in Proposition~\ref{prop:exact_K}.

\vspace{2cm}

\subsection{Eventually single-atom Poisson thinning}\label{sec:Poisson}
In the eventually single-atom thinned-DDP, a prior for the number of atoms discarded by $p_g$, i.e., $u_g$, with $g = 1,2$, is given by setting $u_g - 1 \simind \text{Poisson}(\lambda_g)$. As in the previous section, we analyze the correlation between pairs of random probability measures $p_{1:2}$ by marginalizing with respect to $u_1$ and $u_2$.
\begin{prop}\label{prop:corr_poisson}
Let $p_{1:2}\sim\text{thinned-DDP}(\alpha,P_0,u_{1:2})$ and consider a measurable set $A\in\mathcal{B}$. Assume $u_g - 1 \simind \text{Poisson}(\lambda_g)$, for $g=1,2$. The correlation between $p_1(A)$ and $p_2(A)$ is
\begin{align*}
	&\mathrm{Corr}(p_1(A),p_2(A))  = 
     \mathrm{e}^{-\lambda_1-\lambda_2}\sum_{k=0}^\infty \left(\frac{\alpha}{1+\alpha}\right)^k\left[\left(\frac{\lambda_1}{\lambda_2}\right)^{k/2}+\left(\frac{\lambda_2}{\lambda_1}\right)^{k/2}-\mathbb{I}_{\{0\}}(k)\right]I_k(2\sqrt{\lambda_1\lambda_2}),
\end{align*}
where $I_k$ is the modified Bessel function of the first kind.
\end{prop}
A simplification of the result presented in Proposition~\ref{prop:corr_poisson} arises by directly assigning a Poisson prior to the absolute difference of $u_2 - u_1$, e.g., $|u_2 - u_1| \sim \mathrm{Poisson}(\lambda)$. In this case, the correlation simplifies to  
\begin{equation*}
\text{Corr}(p_1(A), p_2(A)) = \mathrm{e}^{-\lambda / (\alpha + 1)}.
\end{equation*}
As apparent, the correlation approaches one as $\lambda \to 0$ and zero as $ \lambda \to +\infty$. \textcolor{black}{Under the same prior specification for $\ell_{1:2}$, Corollary~\ref{cor:corr_parent_eventually_poisson} in the Supplementary Material presents the correlation between a component of $p_{1:2}$ and the parent process $p$.}\\
As in the previous section, we conduct a numerical analysis to examine how assigning a Poisson prior to the variables $u_g$ in an eventually single-atom thinned-DDP influences the prior distribution of the number of distinct values, both shared across components and component-specific. Our analysis considers the same setting as in Section~\ref{sec:bern}, except that here we assume $p_{1:2} \sim \text{thinned-DDP}(1, P_0, u_{1:2})$, with $u_g - 1 \simiid \mathrm{Poisson}(\lambda)$ for $g = 1,2$. The right panel of Figure~\ref{fig:expected_number_atoms} displays Monte Carlo estimates of $\mathbb{E}[K_0]$, $\mathbb{E}[K_1]$, and $\mathbb{E}[K_2]$ for $n = 100$ and various values of $\lambda$. When $\lambda$ is small, the expected number of covariate-specific clusters, $\mathbb{E}[K_g] $, is close to zero. This occurs because the number of atoms not shared between $p_1$ and $p_2$ coincides with $|u_2 - u_1|$, and since both the distributions of $u_1$ and $u_2$ are concentrated around values close to zero, $|u_2 - u_1|$ has a small mean and variance. As $\lambda$ increases, both the mean and variance of $|u_2 - u_1|$ grow, leading to a larger expected number of unshared atoms and, consequently, an increase in each $\mathbb{E}[K_g]$. Conversely, $\mathbb{E}[K_0]$ follows an opposite trend since the total expected number of clusters, $\mathbb{E}[K_0] + \mathbb{E}[K_g]$, is constant with respect to $\lambda$. The right panel of Figure~\ref{fig:tot_expected_number_atoms} shows the growth of the total expected number of clusters, $\mathbb{E}[K]$, as the sample size $n$ increases. This growth appears more rapid for larger values of $\lambda$. Additionally, as $\lambda$ approaches zero or becomes large, $\mathbb{E}[K]$ gets close to the bounds established in Proposition~\ref{prop:exact_K}.

\section{Thinned-DDP mixture models}
\label{sec::posterior_inference}

In view of the applications in Sections~\ref{sec::simulation} and~\ref{sec::application}, we consider $G$ samples of continuous observations, $y_g = (y_{1,g},\dots,y_{n_g,g})$ for $g = 1,\dots,G$, and summarize the data in the $n$-dimensional vector $y_{1:G} = (y_1,\dots,y_G)$, where $n = n_1+\dots+n_G$. When modeling continuous observations, the almost sure discreteness of the DP makes it unsuitable. A common approach is to convolve it with a continuous kernel. We propose modeling $y_{1:G}$ as partially exchangeable using a Gaussian mixture model with a thinned-DDP vector of mixing measures and Bernoulli priors for the thinning sequences. Namely,
\begin{equation}
    \label{eq:mixture}
\begin{aligned}
y_g\mid f_g &\simind f_g& g=1,\ldots,G \\
f_g&=\phi * p_g & g=1,\ldots,G\\
p_{1:G}&\sim \text{thinned-DDP}(\alpha,P_0,\ell_{1:G})&\\
\ell_{j,g}&\simiid \text{Bern}(\pi_g)&j=1,2,\ldots; \; g=1,\ldots,G
\end{aligned}
\end{equation}
where $\phi(\cdot) = \phi(\cdot \mid \mu, \sigma^2)$ is the density function of a Gaussian random variable with mean $\mu$ and variance $\sigma^2$, and $*$ denotes the convolution operator. 
To ensure conjugacy, we specify the base measure $P_0$ as a normal-inverse gamma distribution with parameters $(\mu_0, \tau_0, \gamma_0, \lambda_0)$. Namely, $(\mu,\sigma^2)\sim P_0$ means that $\mu\mid\sigma^2 \sim \N(\mu_0, \sigma^2/\tau_0)$ and $1/\sigma^2 \sim \text{Gamma} (\gamma_0,\lambda_0)$. The covariate-specific thinning probabilities $\pi_g$ are assigned independent $\text{Beta}(a_{\pi},b_{\pi})$ priors.\\
Although our proposed framework is general, in model~\eqref{eq:mixture} and throughout the remainder of the paper, we focus on the Bernoulli thinning specification introduced in Section~\ref{sec:bern}, as it yields the most straightforward computational scheme.
Given the constructive definition in \eqref{eq::thinnedDDP}, we find it convenient to adapt computational strategies, originally devised for the DP, relying on its stick-breaking representation, to the partially exchangeable setting considered here. Notable examples from the literature include the blocked Gibbs sampler \citep{Ish01}, the slice sampler and its efficient variants \citep{Walker2007,kalli2011slice,Ge05}, and the retrospective sampler \citep{papaspiliopoulos2008retrospective}. These methods, commonly referred to as conditional samplers, operate by explicitly updating the process components within the sampler and rely on either stochastic or deterministic truncation. We outline a blocked Gibbs sampler for the thinned-DDP mixture model with Bernoulli thinning in \eqref{eq:mixture}. The conjugate specification of kernel $\phi$ and base measure $P_0$ allows for the analytical derivation of full conditional distributions for all model parameters. 
The main steps of our algorithm mirror those of a standard blocked Gibbs sampler for a Dirichlet process, with two key distinctions: (i) our approach explicitly updates the thinning sequences $\ell_{1:G}$, and (ii) the update of the common stick-breaking variables $v_j$ accounts for observations across all samples $y_g$, with $g = 1, \dots, G$. The complete algorithm is detailed in Section~\ref{Asec::gibbs_derivation} of the Supplementary Material and its implementation is available in the Github repository \url{https://github.com/laura-dangelo/thinned_DDP}.

\section{Simulation studies}\label{sec::simulation}
We conduct a numerical investigation of the posterior properties of the thinned-DDP mixture model defined in Section~\ref{sec::posterior_inference} using simulated data consisting of $G$ samples. Specifically, we assess the accuracy of the estimated partitions of the observations and that of the posterior density estimates. The study is divided into two parts, presented in Sections~\ref{sec:pool} and~\ref{sec:comp}. First, to illustrate how the thinned-DDP framework enables information sharing across groups while accounting for heterogeneity, we compare its results to two extreme modeling approaches: complete pooling and no pooling. 
In the complete-pooling approach, the data structure is ignored by combining all $G$ samples into a single one and fitting a common DP mixture, thereby maximizing information sharing. This formulation corresponds to setting $\pi_g = 1$ for all groups in the thinned-DDP. 
On the other hand, the no-pooling approach adopts independent DP mixture models for the $G$ samples, thus preventing any information sharing across them. 
Second, to assess the performance of the thinned-DDP, we compare its results with three state-of-the-art modeling approaches: the CAM of \citet{Denti2023}, the GM-DDP of \citet{GMDDP}, and the HDP of \citet{Teh2006}. Like the thinned-DDP, these methods can be used as nonparametric models for grouped data while also inducing a partition of the entire data vector. 
Additionally, the algorithms for performing posterior inference under the CAM and GM-DDP are implemented in the R packages \texttt{sanba} \citep{sanba} and \texttt{BNPmix} \citep{BNPmix}, respectively. 
For the HDP, we used the algorithm of \cite{blockedHDP}, relying on the implementation made available by one of the authors on GitHub.

We simulated grouped data $y_{1:G}$, where each sample $y_g$, with $g=1,\ldots,G$, follows a finite mixture of univariate Gaussians. Specifically, each $y_g$ was drawn from either a three-component mixture with means at $(-5, 0, 5)$ and probabilities $(0.5, 0.25, 0.25)$ or a two-component mixture with means at $(5, 10)$ and probabilities $(0.4, 0.6)$. The component variances were fixed at 0.6 across all mixtures. To evaluate how the number of groups and their sample sizes influence posterior inference, we considered multiple configurations. First, we generated data from $G=2$ populations with varying sample sizes: $(n_1, n_2) \in \{(10,30), (20,60), (30,90), (40,120)\}$. Next, we extended the analysis to $G=10$ populations with sample sizes $(n_{g'}, n_{g''}) \in \{(10,30), (20,60), (30,90), (40,120)\}$ with $g'=1,\dots,5$ and $g''=6,\dots,10$. This setup resulted in total sample sizes $n$ ranging from 40 to 800. In both scenarios, $G=2$ and $G=10$, half of the samples were drawn from the three-component mixture and the other half from the two-component mixture. Each experiment was repeated on 50 independent datasets. The algorithm settings and prior specifications are described in detail in Section~\ref{Asec::simu} of the Supplementary Material.

\subsection{Comparison with complete-pooling and no-pooling mixtures}\label{sec:pool}

\begin{figure}[h]
	\centering
	\includegraphics[width=.8\linewidth]{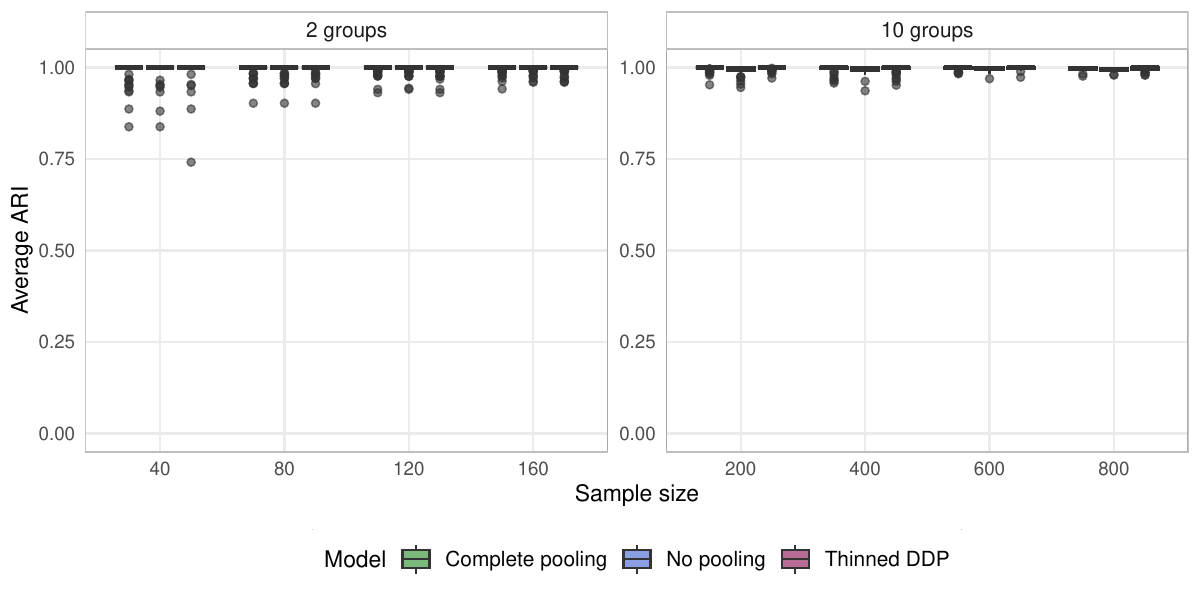}
	\caption{Boxplot of the average ARI for the thinned-DDP, the complete-pooling, and the no-pooling DP mixture on the simulated data. }\label{fig::rand1}
\end{figure}

We begin by evaluating the models' ability to estimate the data partition. To do so, we use the realizations from the posterior distribution of the cluster allocation variables to construct group-specific posterior similarity matrices, where each entry $(i,i')$ represents the posterior probability that observations $i$ and $i'$, with $i, i' \in \{1,\dots,n_g\}$, belong to the same cluster within group $g$. Next, we obtain point estimates of the partitions by minimizing the variation of information loss \citep{wade2018}, implemented via the \texttt{salso} R package \citep{salso, Dahl2022}. To assess accuracy, we compute the Adjusted Rand Index \citep[ARI;][]{rand1971, hubert1985} for each group, measuring the agreement between the estimated and true partitions. As a summary measure, we report the average ARI, calculated as the mean of the group-specific ARIs. 
Figure~\ref{fig::rand1} represents the distribution of the ARI across 50 replications for all combinations of $G$ and $n$. The no-pooling model highlights a key limitation: when the sample size is small, the lack of information sharing between groups substantially impacts performance, leading to a lower ARI in the first configuration. In contrast, both the complete-pooling and thinned-DDP mixtures demonstrate strong performance in estimating the data partition in all settings.

\begin{figure}
	\centering
	\includegraphics[width=.8\linewidth]{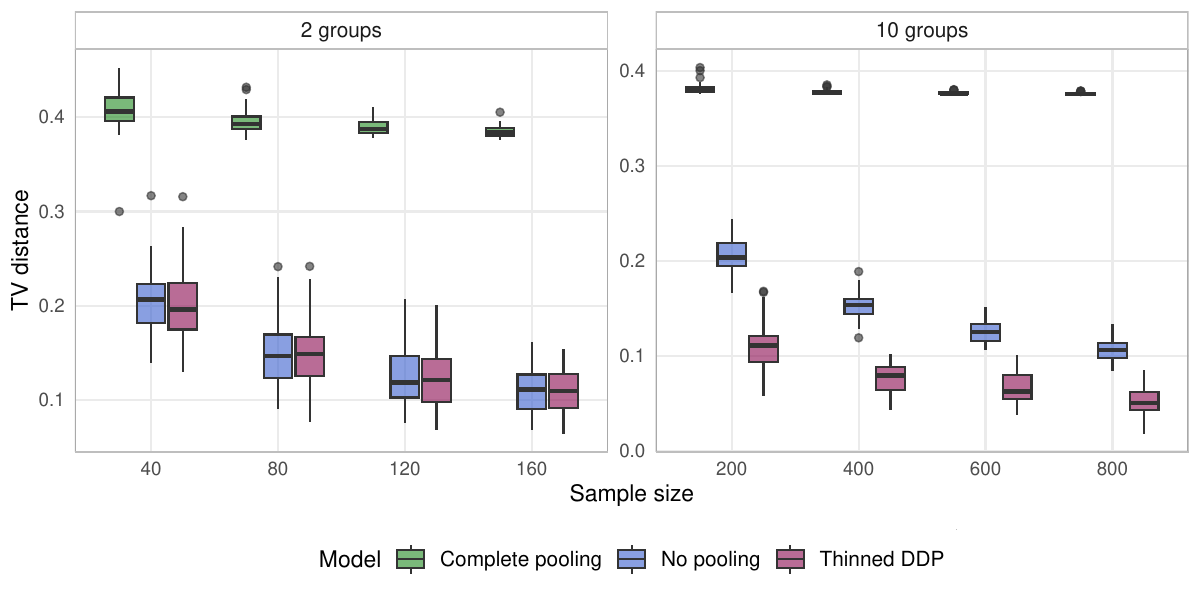}
	\caption{Boxplot of the TV distance for the thinned-DDP, the complete-pooling, and the no-pooling DP mixture on the simulated data. \label{fig::TVdist1}}
\end{figure}

Next, we evaluate the models' accuracy in estimating the data density. At each MCMC iteration $q = 1, \dots, Q$, we computed an estimate of $f_g$ on a grid of 300 equally spaced points $x_i$, which we denote as $\hat{f}^{(q)}_g(x_i)$. The final density estimate at each $x_i$ was then obtained as the Monte Carlo mean $\hat{f}_g(x_i) = \frac{1}{Q} \sum_{q=1}^Q \hat{f}^{(q)}_g(x_i)$. 
To assess the overall accuracy of the posterior point estimate, Figure~\ref{fig::TVdist1} summarizes the distribution of the total variation (TV) distance across scenarios. As expected, the complete pooling approach exhibits the largest discrepancy from the true density in all cases. When $G=2$, the no-pooling and thinned-DDP models demonstrate comparable accuracy. However, for $G=10$, the thinned-DDP consistently achieves a smaller TV distance, particularly for small sample sizes. This improved accuracy in posterior density estimation is likely attributed to its ability to share information across groups. 
To assess the uncertainty of the estimates $\hat{f}_g$ alongside their accuracy, we estimated pointwise 95\% highest posterior density (HPD) credible intervals on the same grid. As an overall measure of uncertainty, we computed the average length of these intervals. 
Figure~\ref{fig::areaCI1} in the Supplementary Material displays a summary of the distribution of this measure. The results clearly show that the thinned-DDP balances between the two extremes of complete and no pooling. The no-pooling model, relying solely on individual samples, exhibits the highest uncertainty, whereas the complete-pooling model, using all data, has the lowest. The thinned-DDP consistently falls between these extremes, demonstrating its ability to borrow information and reduce the variability of group-specific estimates.

\subsection{Comparison with state-of-the-art methods}\label{sec:comp}
\begin{figure}[h]
	\centering
	\includegraphics[width=.8\linewidth]{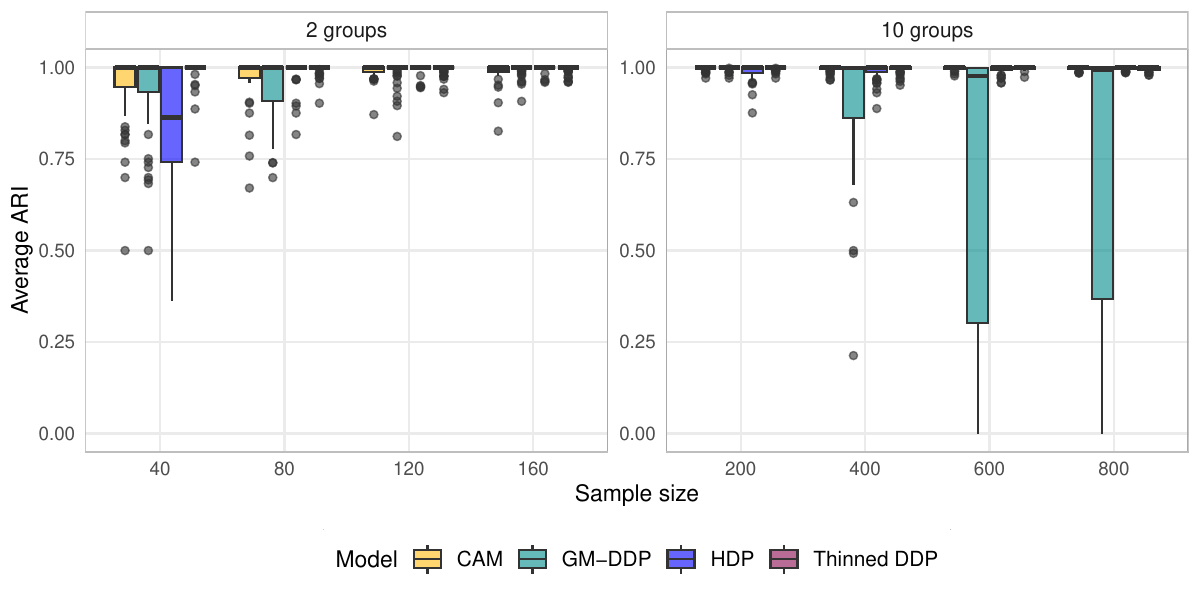}
	\caption{Boxplot of the average ARI for the thinned-DDP, the CAM, the GM-DDP, and the HDP mixture on the simulated data. }\label{fig::rand2}
\end{figure}

We now compare the thinned-DDP with three nonparametric mixture models for dependent grouped data: the CAM, the HDP, and the GM-DDP. As before, we begin by evaluating the models' ability to recover the true data partition.
Figure~\ref{fig::rand2} displays a summary of the distribution of the average ARI, as defined in the previous section, across all replications for each combination of $G$ and sample sizes $n_g$, $g=1,\ldots,G$. The CAM, HDP, and thinned-DDP demonstrate strong performance across scenarios, achieving near-perfect ARI in several configurations. While the GM-DDP performs comparably well when $G=2$, it struggles in settings with a larger number of populations and greater sample sizes. 

\begin{figure}
	\centering
	\includegraphics[width=.8\linewidth]{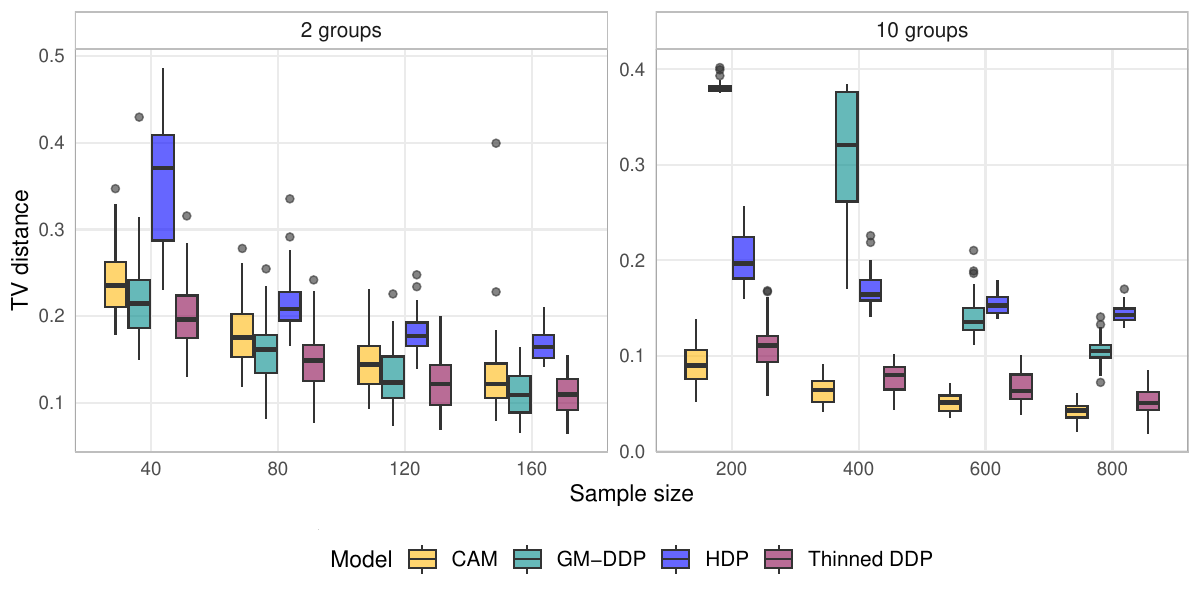}
	\caption{Boxplot of the TV distance for the thinned-DDP, the CAM, the GM-DDP, and the HDP mixture on the simulated data. \label{fig::TVdist2}}
\end{figure}

Turning to the models' ability to estimate the data density, Figure~\ref{fig::TVdist2} provides a summary of the distribution of the TV distance, assessing the accuracy of the posterior point estimates, while Figure~\ref{fig::areaCI2} in the Supplementary Material summarizes the distribution of the average size of the credible bands, evaluating the uncertainty associated with the estimates. 
When $G=2$, the thinned-DDP, CAM, and GM-DDP exhibit similar performance; however, the thinned-DDP provides slightly more accurate and precise density estimates than its competitors. For $G=10$, the CAM's accuracy and precision improve steadily as the sample size increases, yet the thinned-DDP maintains comparable performance across all configurations. In contrast, when $G=10$ and the sample size is small, the GM-DDP appears to struggle with density estimation. 
Finally, the HDP generally yields larger TV distances across the scenarios considered.
To further clarify the behavior of each model and the rationale behind these results, we examine examples of estimated densities and credible bands. Figures~\ref{fig::density_est_2G_minSS}--\ref{fig::density_est_10G_maxSS_hdp} in the Supplementary Material report these quantities for randomly selected datasets generated under different settings.
The graphs compare posterior point estimates and credible bands for the densities estimated by the thinned-DDP, HDP, CAM, and GM-DDP, using the true data-generating density as a reference. A visual inspection highlights that a key advantage of the thinned-DDP over CAM is its ability to prevent excessive information sharing between groups, particularly when populations are heterogeneous and samples are small. In such cases, CAM tends to introduce spurious modes. For $G = 2$, the GM-DDP also effectively avoids this issue. However, when the number of populations increases and sample sizes remain small, the additive structure of the GM-DDP limits its flexibility in capturing heterogeneity. To address this limitation, one could explore more refined formulations, such as those proposed in \citet{galaxySIS}.

\section{Real data analysis}\label{sec::application}
The Collaborative Perinatal Project (CPP) was a large longitudinal study conducted in the United States between 1959 and 1965 to investigate the impact of pregnancy complications on children's health. It enrolled thousands of women from 12 medical centers across the country. Here, we focus on a subset of the data and analyze the gestational age of 2,313 women. While the distributions of this variable across the 12 centers share similar medians and right tails, their left tails differ noticeably. Some hospitals have a considerably higher proportion of patients with very short gestational ages (premature births) than others, suggesting that the hospital sample is not homogeneous  (see Figure~\ref{fig::boxplots} in the Supplementary Material). 
Additionally, the sample sizes of the centers vary widely, ranging from 77 to 481 patients. Collecting data from multiple sources is essential in observational studies, offering significant advantages over single-center analyses, including larger sample sizes, greater population diversity, and improved generalizability. However, within each hospital, unobserved factors can influence patient outcomes, leading to potential heterogeneity among individuals in the same center. An effective model should account for these various sources of variability while flexibly integrating information across samples. The thinned-DDP mixture model is particularly well-suited for this task. Its mixture formulation captures latent heterogeneity within centers, while the presence of shared atoms effectively models similarities in the densities across samples. At the same time, both group-specific atoms and the group-specific weights of shared atoms account for differences, striking a balance between borrowing information and preserving individual center characteristics.

We estimated a thinned-DDP Gaussian mixture model with Bernoulli thinning, as described in Section~\ref{sec::posterior_inference}, by running the blocked Gibbs sampler for 120,000 iterations, discarding the first 20,000 as burn-in and thinning the resulting chains to reduce autocorrelation. 
The truncation parameter was set here equal to 300, the remaining model parameters were set as in Section~\ref{sec::simulation}.
The posterior point estimate of the partition provides a first valuable insight into the data. Thanks to the commonality of the atoms of the thinned-DDP, we can estimate the overall partition, alongside the group-specific ones, to allow for a better characterization of the patients across centers. Minimizing the variation of information loss results in three estimated clusters. The sample means of the observations assigned to the three clusters are 34.1, 39.8, and 44.0 weeks, respectively. Accordingly, we interpret the clusters as subpopulations corresponding to premature births, standard full-term pregnancies, and post-term pregnancies. 

\begin{figure}[t!]
\vspace{.5cm}
	\centering
	\includegraphics[width=.9\linewidth]{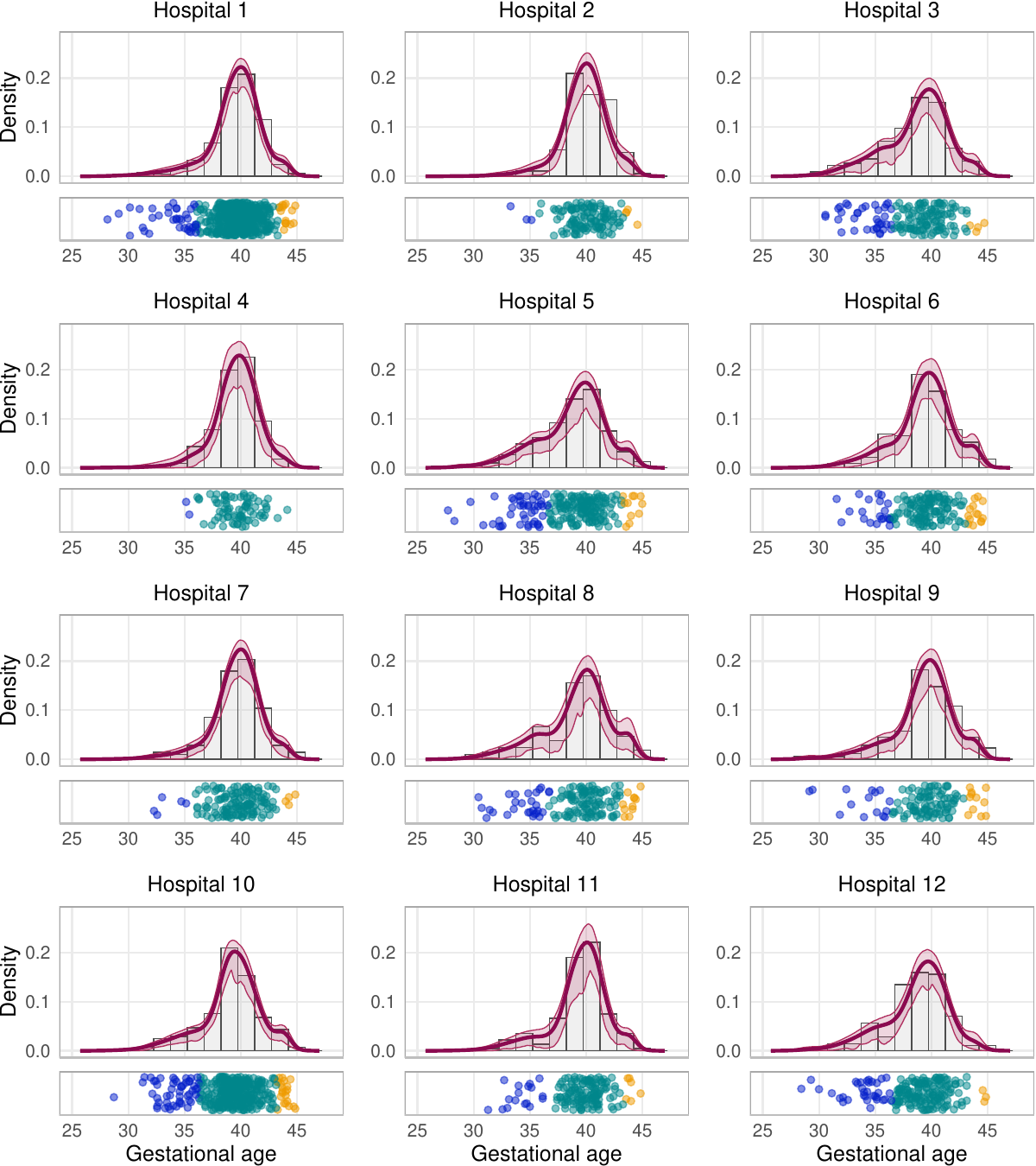}
	\caption{Top panels: histogram of the data, posterior density estimate via thinned-DDP (continuous line) and 95\% pointwise HPD credible intervals (shaded area) for each hospital. Bottom panels: jittered observations, colored by estimated cluster allocation. }\label{fig::densities1}
    \vspace{.5cm}
\end{figure}

In Figure~\ref{fig::densities1}, the bottom panels display the hospital-specific observations colored according to such partition, while the top panels show the histogram of the data, along with the densities' point estimates and 95\% pointwise HPD credible intervals. 
Most observations in all hospitals are assigned to the cluster of standard pregnancies. In contrast, the proportions of pre-term and post-term pregnancies vary substantially across centers. 
For instance, the distribution of Hospital 4 is largely explained by a single cluster, with a second component serving only to account for two outliers. In contrast, Hospital 5 exhibits all three clusters, each containing a substantial number of observations.
The density estimates highlight differences in the relative weight of each subpopulation across hospitals: some densities are approximately symmetric and unimodal, whereas others exhibit clear multimodality and skewness.
This analysis illustrates the flexibility of the thinned-DDP in capturing heterogeneity both within the data, by identifying distinct subpopulations, and across groups, through differences in cluster weights. 
The estimated model and the algorithm's outputs further allow us to assess the similarity between hospital-specific densities. 
For each pair of hospitals, we can track whether they were modeled using the same mixture components over the iterations of the Gibbs sampler. 
Storing this information in a $12\times 12$ matrix returns the frequencies with which each pair of hospitals was modeled by the same mixture density.
This matrix can be interpreted as a posterior similarity matrix and used to perform clustering of the hospitals. Minimizing the variation of information loss, we identified two distinct clusters: the first includes Hospitals 1, 2, 4, and 7, while the second comprises the remaining hospitals. Notably, hospitals in the first cluster are characterized by considerably fewer points assigned to the observational clusters for pre-term and post-term pregnancies. 
This finding aligns with the estimated total variation distance computed across the group-specific densities, as shown in Figure~\ref{fig::TV_CPP} in the Supplementary Material. The plot displays the Monte Carlo estimate of the TV distance between all pairs of hospitals and highlights two distinct blocks of hospitals with similar densities.

\section*{Acknowledgments}
Bernardo Nipoti and Andrea Ongaro are supported by the European Union Next Generation EU funds, component M4C2, investment 1.1, PRIN-PNRR 2022 (P2022H5WZ9).

\section*{Data Availability}
The data that support the findings of this study are openly available in the R package \texttt{BNPmix} \citep{BNPmix} at \url{http://doi.org/10.32614/CRAN.package.BNPmix}.

\bibliographystyle{agsm}
\bibliography{references}

\clearpage

\appendix
\setcounter{figure}{0}
\setcounter{table}{0}
\setcounter{equation}{0}
\renewcommand{\thetable}{\thesection\arabic{table}}
\renewcommand{\thefigure}{\thesection\arabic{figure}}
\renewcommand{\theequation}{\thesection\arabic{equation}}
\counterwithin{cor}{section}

\def\spacingset#1{\renewcommand{\baselinestretch}%
	{#1}\small\normalsize} \spacingset{1}

\spacingset{1.2} 


\begin{center}
	{
		\LARGE
		\vspace{1cm}
		\textbf{\\ Supplementary Material for}\\
		\textbf{``Dependent Dirichlet processes via thinning''\\}
		\vspace{.5cm}
		\large
		\textbf{Laura D'Angelo, Bernardo Nipoti, and Andrea Ongaro }
		\vspace{.4cm}
	}  
\end{center}

\section{Proofs of the results in Section 2}    \label{Asec:conditional_properties}

\subsection{Proof of Proposition~\ref{prop:corr}}\label{Asubsec::conditional_correlation}
We derive the correlation between pairs of distributions distributed as a thinned-DDP.
Since the thinned-DDP belongs to the class of multivariate species sampling processes introduced by \citet{Fra25}, we can specialize their Proposition~2.7 to our setting. Specifically, Proposition~2.7 states that, if $(p_1,p_2)$ are distributed as a multivariate species sampling process, then, for every measurable set $A$ such that $0 < P_0(A) < 1$,
\begin{equation}
    \mathrm{Corr}(p_1(A), p_2(A)) = \frac{\P(y_{i,1} = y_{k,2})}{\sqrt{
    \P(y_{i,1} = y_{i',1}) \P(y_{k,2} = y_{k',2})
    }}.
    \label{Aeq:eq_corr_franz}
\end{equation}
Consider $p_{1:2}\sim\text{thinned-DDP}(\alpha,P_0,\ell_{1:2})$, then $\P(y_{i,g} = \theta_j \mid p_g ) = v_j\ell_{j,g} \prod_{m<j} (1-v_m\ell_{m,g}) $, for $j\geq 1$ and $g=1,2$.
We start by computing the numerator of Eq.~\eqref{Aeq:eq_corr_franz}. 
\begin{align*}
    \P (y_{i,1} = y_{k,2}\mid p_1,p_2) &= \sum_{j\geq 1}  \P (y_{i,1}= j, y_{k,2}=j \mid p_1,p_2) \\
    &=\sum_{j\geq 1}  \P (y_{i,1}= j\mid p_1,p_2)\P (y_{k,2}=j \mid p_1,p_2) \\
     &=\sum_{j\geq 1}  [\ell_{j,1}v_j \prod_{h<j}(1 - \ell_{h,1}v_h) ]  [\ell_{j,2}v_j \prod_{h<j}(1 - \ell_{h,2}v_h) ]\\
    &=\sum_{j\geq 1}  \ell_{j,1}\ell_{j,2}v_j^2 \prod_{h<j}(1 - \ell_{h,1}v_h) (1 - \ell_{h,2}v_h) \\
     &=\sum_{j\geq 1}  \ell_{j,1}\ell_{j,2}v_j^2 \prod_{h<j}\bigg(1 - (\ell_{h,1}+\ell_{h,2})v_h +\ell_{h,1} \ell_{h,2}v_h^2 \bigg) .
\end{align*}
Marginalizing with respect to $\{v_j\}_{j\geq 1}$ we get
\begin{align*}
    \P &(y_{i,1} = y_{k,2}; \ell_{1:2})
     =\sum_{j\geq 1}  \ell_{j,1}\ell_{j,2} \mathbb{E}[v_j^2] \prod_{h<j}\bigg(1 - (\ell_{h,1}+\ell_{h,2}) \mathbb{E}[v_h] +\ell_{h,1} \ell_{h,2}\mathbb{E}[v_h^2] \bigg) \\
&=\sum_{j\geq 1}  \ell_{j,1}\ell_{j,2} \frac{2}{(\alpha + 1)(\alpha+2)} \prod_{h<j}\bigg(1 - (\ell_{h,1}+\ell_{h,2}) \frac{1}{\alpha+1} +\ell_{h,1} \ell_{h,2}\frac{2}{(\alpha + 1)(\alpha+2)}  \bigg)   \\  
&=\frac{2}{(\alpha + 1)(\alpha+2)} \sum_{j\geq 1}  \ell_{j,1}\ell_{j,2}  \prod_{h<j} \frac{1}{(\alpha + 1)(\alpha+2)} \underbrace{\bigg(\alpha^2 + 3\alpha + 2 - (\alpha+2)(\ell_{h,1}+\ell_{h,2}) +2 \ell_{h,1}\ell_{h,2} \bigg) }_{(a_h)}.
\end{align*}
The last term depends on the specific realization of $\ell_{1:2}$. In particular, the quantity within braces is equal to
\begin{equation*} 
(a_h) =
    \begin{cases}
        \alpha(\alpha+2) \text{ if } \ell_{h,1}\neq \ell_{h,2}\\
        (\alpha+1)(\alpha+2)\text{ if } \ell_{h,1}= \ell_{h,2} = 0\\
        \alpha(\alpha+1) \text{ if } \ell_{h,1}= \ell_{h,2} = 1
    \end{cases}
\end{equation*}
which implies that 
\begin{equation*}
    \frac{(a_h)}{(\alpha+1)(\alpha+2)}=\frac{\alpha}{\alpha+\ell_{h,1}+\ell_{h,2}}.
\end{equation*}
So we can write 
\begin{equation*}
    \P (y_{i,1} = y_{k,2} ; \ell_{1:2}) =\frac{2}{(\alpha + 1)(\alpha+2)} \sum_{j\geq 1}  \ell_{j,1}\ell_{j,2}  \prod_{h<j} \frac{\alpha}{\alpha+\ell_{h,1}+\ell_{h,2}}.
\end{equation*}

\noindent Finally, to compute the correlation we need the denominator of~\eqref{Aeq:eq_corr_franz}, which is the product of the probabilities of observing ties within the same group. Since each $p_g$ is distributed as a Dirichlet process, we have
$\P (y_{i,g} = y_{i',g}) = 1/(\alpha + 1)$. \\
Hence the correlation is
\begin{equation*}
     \text{Corr}(p_1(A), p_2(A) ; \ell_{1:2} ) =\frac{2}{(\alpha+2)} \sum_{j\geq 1}  \ell_{j,1}\ell_{j,2}  \prod_{h<j} \frac{\alpha}{\alpha + \ell_{h,1}+\ell_{h,2}}.
\end{equation*}
Denoting with $s_j = \sum_{h=1}^{j-1}\ell_{h,1}\ell_{h,2}$ and $q_j = \sum_{h=1}^{j-1}\mathbb{I}_{\{\ell_{h,1}\neq \ell_{h,2}\}}$, the correlation can be further simplified into
\begin{equation*}
    \mathrm{Corr}\left(p_1(A), p_2(A) ; \ell_{1:2} \right)=\frac{2}{(\alpha+2)} \sum_{j\geq 1}  \ell_{j,1}\ell_{j,2} \left(\frac{\alpha}{\alpha+2} \right)^{s_j} \left(\frac{\alpha}{\alpha+1} \right)^{q_j}.
\end{equation*}

\subsection{Proof of Proposition~\ref{prop:exact_K}}
We let $y_g=\{y_{1,g},\ldots,y_{n_g,g}\}$, with $g=1,2$, and $y=\{y_{1,1},\ldots,y_{n_1,1},y_{1,2},\ldots,y_{n_2,2}\}$. We recall that
\begin{itemize}
    \item $K=K(n_1,n_2;\ell_{1:2})$ is the number of distinct values in $y$;
    \item $K_0=K_0(n_1,n_2;\ell_{1:2})$ is the number of distinct values in $y$ coinciding with observations in both $y_1$ and $y_2$;
    \item $K_g=K_g(n_g;\ell_{1:2})$ is the number of distinct values in $y_g$ that do not coincide with any observation in $y_{3-g}$.
    \end{itemize}
    Moreover, we introduce:
\begin{itemize}
     \item $K_g^{(g)}(n_g;\ell_{1:2})$ is the number of distinct values in $y_g$ corresponding to atoms $\theta_j$ that are specific to $p_g$, that is for which $\ell_{j,g}=1$ and $\ell_{j,3-g}=0$;
     \item $K_0^{(g)}(n_g;\ell_{1:2})$ is the number of distinct values in $y_g$ corresponding to atoms $\theta_j$ appearing in both $p_1$ and $p_2$, that is for which $\ell_{j,1}=\ell_{j,2}=1$;
     \item $K_{0}^{(1:2)}(n_1,n_2;\ell_{1:2})$ is the number of distinct values in $y$  corresponding to atoms $\theta_j$ appearing in both $p_1$ and $p_2$, that is for which $\ell_{j,1}=\ell_{j,2}=1$;
     \item $K^{(g)}(n_g;\ell_{1:2})=K_0^{(g)}(n_g;\ell_{1:2})+K_g^{(g)}(n_g;\ell_{1:2})$ is the number of distinct values in $y_g$. 
\end{itemize} 
We observe that
\begin{equation}\label{eq:fact}K(n_1,n_2;\ell_{1:2})=K^{(1:2)}_{0}(n_1,n_2;\ell_{1:2})+K^{(1)}_1(n_1;\ell_{1:2})+K^{(2)}_2(n_2;\ell_{1:2}),
\end{equation}
almost surely, and that
\begin{equation}\label{eq:fact2}
K^{(1:2)}_{0}(n_1,n_2;\ell_{1:2})=K_0^{(1)}(n_1;\ell_{1:2})+K_0^{(2)}(n_2;\ell_{1:2})-K_0(n_1,n_2;\ell_{1:2}),
\end{equation}
almost surely.
Equation~\ref{eq:fact2} implies that, almost surely,
\begin{equation}\label{eq:ineq1}
    K_{0}^{(1:2)}(n_1,n_2;\ell_{1:2})\leq K^{(1)}_0(n_1;\ell_{1:2})+K^{(2)}_0(n_2;\ell_{1:2}).
\end{equation}
From \eqref{eq:fact} and \eqref{eq:ineq1} we get that, almost surely,
\begin{align*}
K(n_1,n_2;\ell_{1:2})&\leq \sum_{g=1}^2 \left(K_0^{(g)}(n_g;\ell_{1:2})+K_g^{(g)}(n_g;\ell_{1:2})\right)\\
&=\sum_{g=1}^2 K^{(g)}(n_g;\ell_{1:2}).
\end{align*}
Thus, 
\begin{align*}\label{eq:thesis1}
\mathbb{E}[K(n_1,n_2;\ell_{1:2})]
&\leq\sum_{g=1}^2\mathbb{E}[K^{(g)}(n_g;\ell_{1:2})]=\sum_{i=1}^{n_1}\frac{\alpha}{\alpha+i-1}+\sum_{i=1}^{n_2}\frac{\alpha}{\alpha+i-1},
\end{align*}
where the last identity holds as, regardless of $\ell_{1:2}$, 
each $p_g$ is marginally distributed as a DP with concentration parameter $\alpha$ for which we have $\mathbb{E}[K^{(g)}(n_g;\ell_{1:2})]=\sum_{i=1}^{n_g}\alpha/(\alpha+i-1)$ \citep[see][]{Pit06}.\\
Next, we observe that the distribution of $K^{(g)}(n_g;\ell_{1:2})$ is invariant to the specification of $\ell_{1:2}$, as $p_g$ is marginally distributed as a DP with concentration parameter $\alpha$ for any specification of $\ell_{1:2}$. As a consequence, we can write \begin{equation}\label{eq:worst}
    \mathbb{E}[K^{(g)}(n_g;\ell_{1:2})]=\mathbb{E}[K^{(g)}(n_g;\tilde{\ell}_{1:2})]
\end{equation} where $\tilde{\ell}_{1:2}$ is such that the two sequences $\tilde{\ell}_1$ and $\tilde{\ell}_2$ are identical. The same definition of $\tilde{\ell}_{1:2}$ gives rise to the following two almost sure identities:
\begin{align}\label{eq:fact3}
    K_g^{(g)}(n_g;\tilde{\ell}_{1:2})&=0,\\
    K_0^{(1:2)}(n_1,n_2;\tilde{\ell}_{1:2})&=K(n_1,n_2;\tilde{\ell}_{1:2}).\label{eq:fact4}
\end{align}
By combining \eqref{eq:fact} and \eqref{eq:fact2} we get
\begin{align*}
K(n_1,n_2;\ell_{1:2})
&=\sum_{g=1}^2 K_0^{(g)}(n_g;\ell_{1:2})-K_0(n_1,n_2;\ell_{1:2})+\sum_{g=1}^2 K_g^{(g)}(n_g;\ell_{1:2})\\
&=\sum_{g=1}^2 K^{(g)}(n_g;\ell_{1:2})-K_0(n_1,n_2;\ell_{1:2}).
\end{align*}
By taking the expectation of both sides of the last identity, we get
\begin{align*}
    \mathbb{E}[K(n_1,n_2;\ell_{1:2})]&=\sum_{g=1}^2 \mathbb{E}[K^{(g)}(n_g;\ell_{1:2})]-\mathbb{E}[K_0(n_1,n_2;\ell_{1:2})]\\
    &\stackrel{\eqref{eq:worst}}{=}\sum_{g=1}^2 \mathbb{E}[K^{(g)}(n_g;\tilde{\ell}_{1:2})]-\mathbb{E}[K_0(n_1,n_2;\ell_{1:2})]\\
    &
    \geq\sum_{g=1}^2 \mathbb{E}[K^{(g)}(n_g;\tilde{\ell}_{1:2})]-\mathbb{E}[K_0(n_1,n_2;\tilde{\ell}_{1:2})]\\
    &\stackrel{\eqref{eq:fact3}}{=}\sum_{g=1}^2 \mathbb{E}[K_0^{(g)}(n_g;\tilde{\ell}_{1:2})+K_g^{(g)}(n_g;\tilde{\ell}_{1:2})]-\mathbb{E}[K_0(n_1,n_2;\tilde{\ell}_{1:2})]\\
    &=\mathbb{E}\left[\sum_{g=1}^2 K_0^{(g)}(n_g;\tilde{\ell}_{1:2})\right]-\mathbb{E}[K_0(n_1,n_2;\tilde{\ell}_{1:2})]\\
    &\stackrel{\eqref{eq:fact2}}{=}\mathbb{E}[K_0^{(1:2)}(n_1,n_2;\tilde{\ell}_{1:2})+K_0(n_1,n_2;\tilde{\ell}_{1:2})]-\mathbb{E}[K_0(n_1,n_2;\tilde{\ell}_{1:2})]\\
    &\stackrel{\eqref{eq:fact4}}{=}\mathbb{E}[K(n_1,n_2;\tilde{\ell}_{1:2})]\\
    &\stackrel{(a)}{=}\sum_{i=1}^{n_1+n_2}\frac{\alpha}{\alpha+i-1},
\end{align*}
where $(a)$ holds since the specification of $\tilde{\ell}_{1:2}$ implies a complete pooling framework with a single sample $y$ of size $n_1+n_2$ distributed according to a DP with concentration parameter $\alpha$.


\subsection{Proof of Corollary~\ref{corollary:corr_eventually}}
We derive the correlation between $p_1(A)$ and $p_2(A)$, where $p_{1:2}\sim\text{thinned-DDP}(\alpha,P_0,u_{1:2})$. We can specify the correlation in Proposition~1 for this setting by noting that, with this formulation, the first $\min\{ u_1, u_2\}-1$ atoms are discarded from both distributions, the second $\lvert u_2 - u_1 \rvert$ are distribution-specific, while the remaining ones are shared. Since the only components contributing to the correlation are the ones for which both $\ell_{j,1}$ and $\ell_{j,2}$ are non-zero, we are left with a sum for $j\geq \max\{ u_1, u_2\}$.
Moreover, at the generic index j in this set, $s_j = j - \max\{ u_1, u_2\}$ and $q_j = \lvert u_2 - u_1\rvert$. Hence we obtain
\begin{align*}
    \text{Corr}(p_1(A),p_2(A) ; u_1,u_2) &= \frac{2}{\alpha + 2} \left( \frac{\alpha}{\alpha + 1} \right)^{\lvert u_2 - u_1\rvert} \sum_{j \geq \max\{ u_1, u_2\}} \left( \frac{\alpha}{\alpha + 2} \right)^{j - \max\{ u_1, u_2\}} \\
     &= \frac{2}{\alpha + 2} \left( \frac{\alpha}{\alpha + 1} \right)^{\lvert u_2 - u_1\rvert} \frac{\alpha+2}{2} \\&= \left( \frac{\alpha}{\alpha + 1} \right)^{\lvert u_2 - u_1\rvert}. 
\end{align*}

\subsection{Proof of Proposition~\ref{prop:exact_K_eventually}}
Without loss of generality, we assume $u_1=1$ and set $w=u_2-u_1$. Thus all the atoms in $p_2$ are shared by $p_1$, while the first $w$ atoms of $p_1$, that is $\theta_{1:w}=\{\theta_1,\ldots,\theta_{w}\}$, are not shared by $p_2$. We introduce the random variable $B$ consisting of the number of observations in $\{y_{1,1},\ldots,y_{n_1,1}\}$ that coincides with one element of $\theta_{1:w}$. Moreover, we let $K^{1:w}$ be the number of distinct values among these $B$ observations, and $K^{(w+1):\infty}$ the number of distinct values among the remaining $n_1-B+n_2$ observations in $\{y_{1,1},\ldots,y_{n_1,1},y_{1,2},\ldots,y_{n_2,2}\}$ that do not coincide with any of the atoms in $\theta_{1:w}$. We thus have
\begin{equation}\label{eq:split}
        \mathbb{E}[K(n_1,n_2,u_{1:2})]=\mathbb{E}[K^{1:w}]+\mathbb{E}[K^{(w+1):\infty}].
    \end{equation}
We let $v=(v_j)_{j\geq 1}$ and we start by working on the term $\mathbb{E}[K^{1:w}]$, for which we have
\begin{align}\label{eq:firstpart}
  \mathbb{E}[K^{1:w}]&=\mathbb{E}_{v}[\mathbb{E}[K^{1:w}\mid v]]\notag\\\notag
  &=\mathbb{E}_{v}\left[w-\sum_{j=1}^w (1-\omega_j)^{n_1}\right]\\\notag
  &=w-\sum_{j=1}^w\mathbb{E}_{v}\left[ (1-\omega_j)^{n_1}\right]\\\notag
  &=w-\sum_{j=1}^w\mathbb{E}_{v}\left[ \sum_{r=0}^{n_1}\binom{n_1}{r}(-1)^r\omega_j^r\right]\\\notag
  &=w-\sum_{j=1}^w\sum_{r=0}^{n_1}\binom{n_1}{r}(-1)^r\mathbb{E}_{v}\left[\omega_j^r\right]\\\notag
  &\stackrel{(a)}{=}w-\sum_{r=0}^{n_1}\binom{n_1}{r}(-1)^r  \frac{\Gamma(1+r)\Gamma(\alpha)}{\Gamma(1+\alpha+r)} \sum_{j=1}^w \frac{\alpha^j}{(\alpha+r)^{j-1}}\\\notag
  &\stackrel{(b)}{=}\sum_{r=1}^{n_1}(-1)^r\binom{n_1}{r}\frac{\Gamma(1+r)\Gamma(\alpha)}{\Gamma(1+\alpha+r)} \frac{\alpha}{r}\frac{(\alpha+r)^w-\alpha^w}{(\alpha+r)^{w-1}}\\
  &=  \Gamma(\alpha+1)\sum_{r=1}^{n_1}(-1)^{r-1}\binom{n_1}{r} \frac{\Gamma(r)}{\Gamma(\alpha+r)}\left\{1-\left(\frac{\alpha}{\alpha+r}\right)^w\right\},
\end{align}
where (a) holds as
\begin{equation*}
\mathbb{E}_{v}\left[\omega_j^r\right]=\mathbb{E}_{v}\left[v_j^r\prod_{l=1}^{j-1}(1-v_l)^r\right],
\end{equation*}
with $v_j\simiid \mathrm{Beta}(1,\alpha)$ for all $j=1,2,\ldots$; and (b) holds since
\begin{equation*}
    \sum_{j=1}^w \frac{\alpha^{j-1}}{(\alpha+r)^{j-1}}=\begin{cases}
        w& \text{if }r=0,\\
        \frac{1}{r}\frac{(\alpha+r)^w-\alpha^w}{(\alpha+r)^{w-1}}& \text{if }r\geq 1.
    \end{cases}
\end{equation*}
Moreover, we observe that
\begin{align}\label{eq:secondpart}
    \mathbb{E}[K^{(w+1):\infty}]&=\mathbb{E}_{v}[\mathbb{E}_{B\mid v}[\mathbb{E}[K^{(w+1):\infty}\mid B,v]]]\notag\\
    &=\mathbb{E}_{v}\left[\sum_{m=0}^{n_1}\mathbb{E}[K^{(w+1):\infty}\mid B=m,v]\Pr(B=m\mid v)\right]\notag\\
    &\stackrel{(c)}{=}\sum_{m=0}^{n_1}\mathbb{E}_{v}\left[\mathbb{E}[K^{(w+1):\infty}\mid B=m,v]\right]\mathbb{E}_{v}\left[\Pr(B=m\mid v)\right],
\end{align}
where (c) holds as the distribution of $B$ only depends on the first $w$ elements of the sequence $v$, say $v_{1:w}$, while, conditionally on $B=m$, the distribution of $K^{(w+1):\infty}$ only depends on the remaining elements of the sequence $v$, say $v_{(w+1):\infty}$. Moreover, we have 
\begin{align}\label{eq:probBm}
   \mathbb{E}_{v}\left[\Pr(B=m\mid v)\right]&=\mathbb{E}_{v_{1:w}}\left[\Pr(B=m\mid v_{1:w})\right] \notag\\
   &= \mathbb{E}_{v_{1:w}}\left[\binom{n_1}{m}\left(\sum_{j=1}^w\omega_j\right)^{m}\left(1-\sum_{j=1}^w\omega_j\right)^{n_1-m}\right]\notag\\
   &=\binom{n_1}{m}\mathbb{E}_{v_{1:w}}\left[\left(1-\prod_{j=1}^w(1-v_j)\right)^{m}\left(\prod_{j=1}^w (1- v_j)\right)^{n_1-m}\right]\notag\\
   &=\binom{n_1}{m}\mathbb{E}_{v_{1:w}}\left[\sum_{l=0}^m(-1)^l\binom{m}{l}\prod_{j=1}^w(1-v_j)^{l}\prod_{j=1}^w (1-v_j)^{n_1-m}\right]\notag\\
   &=\binom{n_1}{m}\sum_{l=0}^m(-1)^l\binom{m}{l}\prod_{j=1}^w\mathbb{E}_{v_j}\left[(1-v_j)^{l+n_1-m}\right]\notag\\
     &=\binom{n_1}{m}\sum_{l=0}^m(-1)^l\binom{m}{l}\left(\frac{\alpha}{\alpha+l+n_1-m}\right)^w,
\end{align}
and
\begin{align}\label{eq:thirdpart}
    \mathbb{E}_{v}\left[\mathbb{E}[K^{(w+1):\infty}\mid B=m,v]\right]
    &= \mathbb{E}\left[K^{(w+1):\infty}\mid B=m\right]\notag\\
    &=\sum_{i=1}^{n_2+n_1-m}\frac{\alpha}{\alpha+i-1},
\end{align}
as, conditionally on $B=m$, $\mathbb{E}[K^{(w+1):\infty}]$ is the expected value of the number of distinct values in a sample of dimension $n_2+n_1-m$ distributed as a DP with concentration parameter $\alpha$ \citep[see][]{Pit06}. By combining \eqref{eq:secondpart}, \eqref{eq:probBm} and \eqref{eq:thirdpart}, we get
\begin{align}\label{eq:lastpart}
    \mathbb{E}[K^{(w+1):\infty}]&=\sum_{m=0}^{n_1}\left(\sum_{i=1}^{n_2+n_1-m}\frac{\alpha}{\alpha+i-1}\right)\binom{n_1}{m}\sum_{l=0}^m (-1)^l \binom{m}{l} \left(\frac{\alpha}{\alpha+l+n_1-m}\right)^w\notag\\ &=\sum_{m=0}^{n_1}\binom{n_1}{m}\left(\sum_{i=1}^{n_2+n_1-m}\frac{\alpha}{\alpha+i-1}\right)\sum_{l=0}^m (-1)^l \binom{m}{l} \left(\frac{\alpha}{\alpha+l+n_1-m}\right)^w.
\end{align}
The proof is completed by replacing the expressions in \eqref{eq:firstpart} and \eqref{eq:lastpart} in \eqref{eq:split}. We observe that the resulting expression does not depend on the assumption that $u_1=1$, which thus can be relaxed.
\setcounter{equation}{0}
\section{Proofs of the results in Section 3} \label{Asec:random_thinning}

\subsection{Proof of Proposition~\ref{prop::support}} \label{Asubsec::support}
We want to prove that the weak support of the thinned-DDP is $\mathcal{P}(\Theta)^G$.
It is simple to check that $\mathcal{P}(\Theta)^G$ is a closed set with probability measure one (see the proof of Theorem~1 in \citeauthor{Barrientos2012},~\citeyear{Barrientos2012}). It is left to show that $\mathcal{P}(\Theta)^G$ is the smallest closed set with probability one. In the proof of Theorem~1, \citet{Barrientos2012} demonstrate that a general sufficient condition for this is to ensure that the process assigns positive probability to certain basic open sets of $\mathcal{P}(\Theta)^G$, which can be written as a product space of particular subsets of simplices. In the following, we show that such a condition holds for the thinned-DDP.

We study the setting where we only have two groups, but the proof can be easily generalized to any $G$. 
Specifically, the sufficient condition  of \citet{Barrientos2012} (see their formula (3)) can be reformulated as
\begin{equation}
	\mathbb{P}\left( \lvert p_g(A_j) - q_{j,g} \rvert <\epsilon, \, j=1,\dots,k, \, g=1,2 \right) = \mathbb{P}(\Omega_0) >0,
	\label{eq:condition_support}
\end{equation}
where $A_1,\dots,A_k$ is a $P_0$-measurable partition of $\Theta$, and the $q_{j,g}$ are constants such that $q_{j,g}\geq 0$, $\sum_{j=1}^{k}q_{j,g}=1$ for $g=1,2$. Furthermore,  we  only have to consider sets $A_j$'s such that $P_0(A_j)>0$ for $j=1,\dots,k$, since if $P_0(A_j)=0$ then $p_g(A_j) = 0$ a.s. and $q_{j,g}=0$. 
We need to find a subset $\Omega_0$ with positive probability.
%
Consider the set
\begin{equation}
    \begin{aligned}
 \Omega_1 = \{ &\ell_{j,1} = 1 \: \text{for $j=1,\dots,k$},
 \\& 
\ell_{j,2} = 0 \: \text{for $j=1,\dots,k$ and }  \ell_{j,2} = 1 \: \text{for $j=k+1,\dots,2k$} \},
\end{aligned} 
\label{supp::eq::omega1}
\end{equation}
while $\ell_{j,1} \in \{0,1\}$ for $j>k$, and $\ell_{j,2} \in \{0,1\}$ for $j>2k$, without restrictions.

By assumption $\mathbb{P}(\Omega_1)>0$.
\noindent Consider now 
\begin{equation}
	\Omega_2 = \{ \omega: \theta_j \in A_j \text{ for $j=1,\dots,k,$ and } \theta_{k+j} \in A_j \text{ for $j=1,\dots,k$} \}.
 \label{supp::eq::omega2}
\end{equation}
Since $P_0(A_j)>0$ for all $j$ and the $\theta_j$ are independent, we have $\mathbb{P}(\Omega_2)>0$.

\noindent Notice now that on $\Omega_1$ we have
\begin{equation*}
	p_1(A_j) = \sum_{j=1}^{\infty} w_{j,1} \delta_{\theta_{j}}(A_j)
\end{equation*}
with 
\begin{gather*}
	w_{j,1} = v_j \prod_{h<j}(1-v_h) \:\text{ for $j=1,\dots,k$}\\
	w_{j,1} = v_j\ell_{j,1} \prod_{h<j}(1-\ell_{h,1}v_h) \:\text{ for $j>k$}
\end{gather*}
and
\begin{equation*}
	p_2(A_j) = \sum_{j=1}^{\infty} w_{j,2} \delta_{\theta_{j}}(A_j)
\end{equation*}
with 
\begin{gather*}
	w_{j,2} = 0 \:\text{ for $j=1,\dots,k$}\\
	w_{j,2} = v_j \prod_{h=k+1}^{j-1}(1-v_h) \:\text{ for $j= k+1,\dots,2k$}\\
 w_{j,2} = v_j\ell_{j,2} \prod_{h<j}(1-\ell_{h,2}v_h) \:\text{ for $j>2k$}.
\end{gather*}
Thus, $(w_{1,1}, \dots,w_{k,1})=\bm{w}_k$ and $(w_{k+1,2}, \dots,w_{2k,2})=\bm{w}'_k$ are independent, given $\Omega_1$.

\noindent Consider now
\begin{equation}
    \Omega_3 = \{ 
    \omega: q_{j,1} - \frac{\epsilon}{k+1} < w_{j,1} < q_{j,1} + \frac{\epsilon}{k+1} ,\: j=1,\dots,k
    \}
    \label{supp::eq::omega3}
\end{equation}
and
\begin{equation*}
    S^k = \{ x_j \in \mathbb{R}^k: x_j>0, \sum_{j=1}^k x_j <1
    \}.
\end{equation*}
Then, $\mathbb{P}(\Omega_3\mid\Omega_1) = \mathbb{P} (\bm{w}_k \in B_3\mid \Omega_1)$ where
\begin{equation*}
    B_3 =\{ \bm{y}_k \in S^k: q_{j,1} - \frac{\epsilon}{k+1} < y_{j,1} < q_{j,1} + \frac{\epsilon}{k+1}, \: j=1,\dots,k \}.
\end{equation*}
$B_3$ is an open subset of $S^k$. 
On $\Omega_1$ we have $\bm{w}_k = f(v_1,\dots,v_k)$, with $w_{1,1} = v_1$, $w_{j,1} = v_j\prod_{h<j}(1-v_h)$: this mapping is one-to-one and continuous from $(0,1)^k$ to $S^k$. It follows that $f^{-1}(B_3) = C_3$ is an open subset of $(0,1)^k$. Therefore,
\begin{equation*}
    \mathbb{P}(\Omega_3\mid\Omega_1) = \mathbb{P}(\bm{w}_k \in B_3\mid \Omega_1) = \mathbb{P}((v_1,\dots,v_k) \in C_3) >0
\end{equation*}
since $(v_1,\dots,v_k)$ has positive density (w.r.t. the Lebesgue measure) on $(0,1)^k$.

A similar argument holds for $\bm{w}_k'$. 
Define 
\begin{equation}
    \Omega_4 = \{ \omega: q_{j,2} - \frac{\epsilon}{k+1} < w_{k+j,2} <   q_{j,2} + \frac{\epsilon}{k+1},  \: j=1,\dots,k\}.
    \label{supp::eq::omega4}
\end{equation}
Then, $\mathbb{P}(\Omega_4\mid\Omega_1) = \mathbb{P} (\bm{w}'_k \in B_4\mid \Omega_1)$ where
\begin{equation*}
    B_4 =\{ \bm{y}'_k \in S^k: q_{j,2} - \frac{\epsilon}{k+1} < y_{j,2} < q_{j,2} + \frac{\epsilon}{k+1}, \: j=1,\dots,k \}.
\end{equation*}
On $\Omega_1$ we have $\bm{w}'_k = f(v_{k+1},\dots,v_{2k})$, with $f$ as before. Therefore $f^{-1}(B_4) = C_4$ is an open subset of $(0,1)^k$ and
\begin{equation*}
    \mathbb{P}(\Omega_4\mid\Omega_1) = \mathbb{P}(\bm{w}_k \in B_4\mid \Omega_1) = \mathbb{P}((v_{k+1},\dots,v_{2k}) \in C_4) >0.
\end{equation*}
Furthermore, 
\begin{equation*}
\mathbb{P}(\Omega_1 \cap \Omega_3 \cap \Omega_4) = \mathbb{P}(\Omega_3, \Omega_4 \mid \Omega_1)\mathbb{P}(\Omega_1)
= \mathbb{P}(\Omega_3 \mid \Omega_1) \mathbb{P}(\Omega_4 \mid \Omega_1) \mathbb{P}(\Omega_1)
\end{equation*}
since $\Omega_3$ and $\Omega_4$ are independent given $\Omega_1$. \\

\noindent Consider now $\Omega_5 = (\Omega_1 \cap \Omega_2 \cap \Omega_3 \cap \Omega_4)$:
$$\mathbb{P}(\Omega_5) = \mathbb{P}(\Omega_1 \cap \Omega_3 \cap \Omega_4) \mathbb{P}(\Omega_2) >0$$ since  $\{\theta_j\}$, $\{\ell_{j,g}\}$, and $\{v_j\}$ are independent.\\

\noindent Let us now show that $\Omega_5 \subseteq \Omega_0$, i.e., that conditions (\ref{supp::eq::omega1}--\ref{supp::eq::omega4}) imposed on $\{\theta_j\}$, $\{\ell_{j,g}\}$, and $\{v_j\}$ imply that $\lvert p_g(A_j) - q_{j,g} \rvert <\epsilon, \, j=1,\dots,k, \, g=1,2$. This completes the proof since then $\mathbb{P}(\Omega_0) \geq \mathbb{P}(\Omega_5)>0$.\\

\noindent Inequalities in $\Omega_3$ imply that 
$$q_{j,1} - \frac{\epsilon}{k+1} < w_{j,1},\quad j=1,\dots,k.$$
Summing w.r.t. $j$ we get
$$1 - \frac{k}{k+1}\epsilon < \sum_{j=1}^k w_{j,1}$$
or,
$$ \sum_{j>k} w_{j,1} < \epsilon \frac{k}{k+1}. $$

\noindent On $\Omega_5$ we have, for $j=1,\dots,k$
\begin{align*}
    q_{j,1} - \frac{\epsilon}{k+1} < w_{j,1} \leq p_1(A_j) &\leq w_{j,1} + \sum_{h>k} w_{h,1} \\ &< q_{j,1} + \frac{\epsilon}{k+1}  + \frac{\epsilon k}{k+1} = q_{j,1} +\epsilon
\end{align*}
since on $\Omega_5$ 
$$ p_1(A_j) = w_{j,1} + \sum_{h>k: \theta_h\in A_j} w_{h,1}. $$

\noindent Similarly, for $p_2$, on $\Omega_5$ we have, for $j=1,\dots,k$
\begin{equation*}
p_2(A_j) = w_{k+j,2} + \sum_{h>2k: \theta_h\in A_{j}} w_{h,2}. 
\end{equation*}

\noindent As before, conditions in $\Omega_4$ imply 
\begin{equation*}
    1- \epsilon \frac{k}{k+1} < \sum_{j = k+1}^{2k} w_{j,2}
\end{equation*}
and 
\begin{equation*}
    \sum_{j >2k} w_{j,2} < \epsilon \frac{k}{k+1} .
\end{equation*}

\noindent Again on $\Omega_5$ we have for $j=1,\dots,k$
\begin{align*}
    q_{j,2} - \frac{\epsilon}{k+1} < w_{k+j,2} \leq p_2(A_j) &\leq w_{k+j,2} + \sum_{h>2k} w_{h,2} \\& < q_{j,2} + \frac{\epsilon}{k+1}  + \frac{\epsilon k}{k+1} = q_{j,2} +\epsilon,
\end{align*}
therefore, $\Omega_5 \subseteq \Omega_0$ and the proof is completed.


\subsection{Proof of Proposition~\ref{prop:corr_bernoulli}}
Suppose $\{\ell_{j,1}\}_{j\geq 1}\simiid \text{Bern}(p)$, $\{\ell_{j,2}\}_{j\geq 1}\simiid \text{Bern}(q)$, and that the sequences are independent. 
The covariance (marginalizing w.r.t. $\ell_{1:2}$) is
\begin{align*}
	\text{Cov}(p_1(A),p_2(A)) &= 
    \mathbb{E}\big[ (p_1(A) -\mathbb{E}[p_1(A)] ) (p_2(A) -\mathbb{E}[p_2(A)] )  \big]\\
    &= \mathbb{E}[p_1(A)p_2(A)] - P_0(A)^2\\
    &=\mathbb{E}[\mathcal{E}_1] + \mathbb{E}[\mathcal{E}_2] + \mathbb{E}[\mathcal{E}_3] - P_0(A)^2
\end{align*}
where
\begin{equation*}
	\mathcal{E}_{1} = \E \left[ \sum_{h=1}^{\infty} \sum_{j=h+1}^{\infty} \ell_{h,1}\ell_{j,2} v_j v_h \prod_{i=1}^{h-1}(1-\ell_{i,1}v_i) \prod_{k=1}^{j-1}(1-\ell_{k,2}v_k) \delta_{\theta_j}\delta_{\theta_h} \right] 
\end{equation*}
\begin{equation*}
	\mathcal{E}_{2} = \E \left[ \sum_{h=1}^{\infty} \sum_{j=h+1}^{\infty} \ell_{h,2}\ell_{j,1} v_j v_h \prod_{i=1}^{h-1}(1-\ell_{i,2}v_i) \prod_{k=1}^{j-1}(1-\ell_{k,1}v_k) \delta_{\theta_j}\delta_{\theta_h} \right] 
\end{equation*}
\begin{equation*}
	\mathcal{E}_{3} = \E\bigg[ \sum_{j=1}^{\infty} \ell_{j,1}\ell_{j,2} v_j^2 \prod_{i=1}^{j-1}(1-\ell_{i,1}v_i) \prod_{k=1}^{j-1}(1-\ell_{k,2}v_k) \delta_{\theta_j}  \bigg]
\end{equation*}

\begin{align*}
	\mathbb{E}& [\mathcal{E}_1] =
	\mathbb{E}\left[ \frac{P_0(A)^2}{(\alpha+1)^2(\alpha+2)}\sum_{j_1=1}^{\infty}\ell_{j_1,1}(\alpha+2(1-\ell_{j_1,2}))\prod_{i_1=1}^{j_1-1}\frac{\alpha}{\alpha+\ell_{i_1,1}+\ell_{i_1,2}}\sum_{j_2=j_1+1}\ell_{j_2,2}\prod_{i_2=j_1+1}^{j_2-1}\frac{\alpha}{\alpha+\ell_{i_2,2}} \right] \\
	&= \frac{P_0(A)^2}{(\alpha+1)^2(\alpha+2)}\sum_{j_1=1}^{\infty} p (\alpha+2(1-q))\prod_{i_1=1}^{j_1-1} \frac{\alpha^2 + \alpha(3-p-q) + 2(1+pq-p-q)}{(\alpha+1)(\alpha+2)}\\ 
	&\quad\times
	\sum_{j_2=j_1+1} q\prod_{i_2=j_1+1}^{j_2-1} \frac{\alpha + 1-q}{\alpha + 1} \\
	&= \frac{P_0(A)^2 p q (\alpha+2(1-q))}{(\alpha+1)^2(\alpha+2)} \sum_{j_1=1}^{\infty}  \left( \frac{\alpha^2 + \alpha(3-p-q) + 2(1+p q-p-q)}{(\alpha+1)(\alpha+2)}\right)^{j_1-1}\hspace{-.25cm} \sum_{j_2=j_1+1}\hspace{-.2cm} \left(\frac{\alpha + 1-q}{\alpha + 1} \right)^{j_2-j_1-1}\\
	&= P_0(A)^2 \frac{p(\alpha + 2 - 2q) }{\alpha(p+q) + 2(p+q-p q)}
\end{align*}

\noindent Similarly, $\mathbb{E}[\mathcal{E}_2] = P_0(A)^2 \frac{q(\alpha + 2 - 2p) }{\alpha(p+q) + 2(p+q-p q)}$.
\begin{align*}
	\mathbb{E}[\mathcal{E}_3] 
	&= \mathbb{E}\left[ \frac{2P_0(A)}{(\alpha+1)(\alpha+2)} \sum_{j=1}^\infty \ell_{j,1}\ell_{j,2}\prod_{i=1}^{j-1}\frac{\alpha}{\alpha+\ell_{i,1}+\ell_{i,2}} \right] \\
	& =\frac{2P_0(A)}{(\alpha+1)(\alpha+2)} p q   \sum_{j=1}^\infty \left(\frac{\alpha^2 + \alpha(3-p-q) + 2(1+pq-p-q)}{(\alpha+1)(\alpha+2)} \right)^{j-1}\\
	& =\frac{2\,p \,q\, P_0(A)}{\alpha(p+q)+2(p+q-p q)}  
\end{align*}
Hence,
\begin{align*}
	\text{Cov}(p_1(A),p_2(A)) 
	&=  \frac{P_0(A)^2 [p(\alpha + 2 - 2q) + q(\alpha + 2 - 2p)] }{\alpha(p+q) + 2(p+q-p q)}  +
	\frac{2\,p \,q\, P_0(A)}{\alpha(p+q)+2(p+q-p q)} -P_0(A)^2\\
	&= \frac{-2 p\, q\, P_0(A)^2}{\alpha(p+q)+2(p+q-p q)} + \frac{2p\,q\,P_0(A)}{\alpha(p+q)+2(p+q-p q)}\\
	&= \frac{2 p\, q\, }{\alpha(p+q)+2(p+q-p q)} P_0(A) \big( 1- P_0(A) \big)
\end{align*}
Finally, the correlation is obtained as 
\begin{align*}
	\text{Corr}(p_1(A),p_2(A)) 
	&= \frac{2\, p\,q (\alpha+1) }{\alpha(p+q)+2(p+q-p q)} 
\end{align*}

\subsection{Proof of Proposition~\ref{prop:corr_poisson}}
Consider again Equation~\eqref{Aeq:eq_corr_franz} from \citet{Fra25}.
The marginal probability of ties can be obtained by marginalizing $u_1$ and $u_2$ as
\begin{align*}
    \P(y_{i,1} &= y_{k,2}) = \mathbb{E}_{u_1,u_2}\big[\P(y_{i,1} = y_{k,2}\mid u_1,u_2)\big]\\
    &\stackrel{(a)}{=} \mathbb{E}_{u_1,u_2} \left[ \mathrm{Corr}(p_1(A), p_2(A)\mid u_1,u_2) \sqrt{    \P(y_{i,1} = y_{i',1}\mid u_1,u_2) \P(y_{k,2} = y_{k',2}\mid u_1,u_2) }    \right]\\
    &\stackrel{(b)}{=}\mathbb{E}_{u_1,u_2} \left[ \mathrm{Corr}(p_1(A), p_2(A)\mid u_1,u_2) \sqrt{    \P(y_{i,1} = y_{i',1}) \P(y_{k,2} = y_{k',2}) }    \right]\\
    &=\mathbb{E}_{u_1,u_2} \left[ \mathrm{Corr}(p_1(A), p_2(A)\mid u_1,u_2)    \right] \sqrt{    \P(y_{i,1} = y_{i',1}) \P(y_{k,2} = y_{k',2}) } ,
\end{align*}
where $(a)$ follows the application of Eq.~\eqref{Aeq:eq_corr_franz} to the multivariate species sampling process conditioned on $(u_1,u_2)$, and $(b)$ follows the fact that the probability of ties within each distribution does not depend on the specific values of $u_1$ and $u_2$. Hence, 
\begin{align*}
    \mathrm{Corr}(p_1(A), p_2(A)) &= \frac{\mathbb{E}_{u_1,u_2} \left[ \mathrm{Corr}(p_1(A), p_2(A)\mid u_1,u_2)    \right] \sqrt{    \P(y_{i,1} = y_{i',1}) \P(y_{k,2} = y_{k',2}) }}{\sqrt{
    \P(y_{i,1} = y_{i',1}) \P(y_{k,2} = y_{k',2})
    }}\\
    &= \mathbb{E}_{u_1,u_2} \left[ \mathrm{Corr}(p_1(A), p_2(A)\mid u_1,u_2)    \right].
\end{align*}

Consider the result of Corollary~1. 
If $(u_1-1)$ and $(u_2-1)$ have independent Poisson distributions with parameters $\lambda_1$ and $\lambda_2$, then $u_2-u_1$ has a Skellam distribution and 
	\begin{align*}
		\text{Corr}(p_1(A),p_2(A)) = \text{e}^{-\lambda_1-\lambda_2}\left(I_0(2\sqrt{\lambda_1\lambda_2})+\sum_{k=1}^\infty \left(\frac{\alpha}{1+\alpha}\right)^k\left[\left(\frac{\lambda_1}{\lambda_2}\right)^{k/2}+\left(\frac{\lambda_2}{\lambda_1}\right)^{k/2}\right]I_k(2\sqrt{\lambda_1\lambda_2})\right)
	\end{align*}

If $\lvert u_2-u_1 \rvert \sim \mathrm{Poisson}(\lambda)$, we compute the marginal correlation starting from the conditional one. Denote $\lvert u_2-u_1 \rvert = x$, then
\begin{align*}
\text{Corr}(p_1,p_2) &= \mathbb{E}_x \left[\left(\frac{\alpha}{\alpha+1}\right)^x \right]\\
&= \sum_{x=0}^{\infty}\left(\frac{\alpha}{\alpha+1}\right)^x e^{-\lambda} \frac{\lambda^x}{x!}\\
&= e^{-\lambda} \sum_{x=0}^{\infty}\left(\frac{\alpha\lambda}{\alpha+1}\right)^x \frac{1}{x!} e^{-\frac{\alpha\lambda}{\alpha+1}} e^{\frac{\alpha\lambda}{\alpha+1}}\\
& = \exp\left\{-\lambda + \frac{\alpha\lambda}{\alpha+1}\right\} =  \exp \left\{\frac{-\alpha\lambda - \lambda + \alpha\lambda}{\alpha+1} \right\}  = \exp \left\{\frac{-\lambda}{\alpha+1} \right\}.
\end{align*}

\setcounter{equation}{0}
\section{Other prior distributions for the thinning variables} \label{Asec:additional_priors}

In this section, we outline two additional prior distributions for the thinning sequences that ensure full weak support of the model while allowing more refined specifications of the dependence between random measures.

\subsection{Dependent-Bernoulli thinning}\label{subsec:dependent_bernoulli}
The prior specification in Section~\ref{sec:bern} assumed independent Bernoulli random variables $\ell_{j,g}\sim\mathrm{Bern}(\pi_g)$ for all group $g$ and index $j$. With this prior, we are only modeling the \textit{marginal} probability of observing atom $j$ in distribution $p_g$, regardless of the presence or absence of the atom in the other distributions. A more refined specification could instead explicitly model this dependence.

Consider, for simplicity, two distributions $p_1$ and $p_2$. We introduce a dependent prior for $\ell_{1:2}$. Specifically, for all $j\geq 1$, we model the pair $(\ell_{j,1},\ell_{j,2})$ as dependent Bernoulli random variables with probabilities:
\begin{equation*}
    \begin{cases}
    \pi_{11} = \mathrm{Pr}(\ell_{j,1} = \ell_{j,2} = 1)\\
    \pi_{01} = \mathrm{Pr}(\ell_{j,1}=0, \ell_{j,2} = 1)\\
    \pi_{10} = \mathrm{Pr}(\ell_{j,1}=1, \ell_{j,2} = 0)\\
    \pi_{00} = \mathrm{Pr}(\ell_{j,1} = \ell_{j,2} = 0)
    \end{cases}
\end{equation*}
under the constraint that $\pi_{11} + \pi_{00} + \pi_{01} + \pi_{10} = 1$. The pairs are still independent for different $j\geq 1$. 

Similar to the other prior specifications, it is possible to compute the marginal correlation between $p_1$ and $p_2$, marginalizing the sequences $\ell_{1:2}$. 
The marginal correlation is then obtained as
\begin{equation*}
    \mathrm{Corr}(p_1,p_2) =     \frac{2 \pi_{11}(\alpha+1)}{(\alpha + 2) (1-\pi_{00}) + \alpha \pi_{11}}  
\end{equation*}

An alternative parameterization allows modeling, instead of all the joint probabilities, the marginal probabilities and the correlations between all pairs. With this parameterization, it is intuitive how to extend the framework to more than two groups.

\subsection{Symmetric eventually single-atom thinning}\label{subsec:symmetric_ev}
The eventually single-atom formulation defined in Section~2.1 implies a nested structure between distributions. However, it is straightforward to relax the constraint by defining a symmetric counterpart. Consider, for simplicity, the case with two distributions. 
We define a blocked structure for the thinning sequences, so that a set of distribution-specific atoms characterizes each random measure. For example, one could define the following block structure:  
\begin{itemize}
	\item $\ell_{j,1} = 1$ and $\ell_{j,2} = 1$ for $j=1,\dots,b_0$,
	\item $\ell_{j,1} = 1$ and $\ell_{j,2} = 0$ for $j=b_0+1,\dots,b_0+b_1$,
	\item $\ell_{j,1} = 0$ and $\ell_{j,2} = 1$ for $j=b_0+b_1+1,\dots,b_0+b_1+b_2$,
	\item $\ell_{j,1} = 1$ and $\ell_{j,2} = 1$ for $j\geq b_0+b_1+b_2+1$.
\end{itemize}
Hence, the distributions share the first $b_0$ atoms; the following $b_1$ atoms are specific to $p_1$; then, $b_2$ are specific to $p_2$; and, finally, all remaining atoms are common:
\begin{align*}
	&p_1 (\cdot)= \sum_{j=1}^{b_0+b_1}\left[ v_j \prod_{h=1}^{j-1} (1-v_h)\right]\delta_{\theta_j} (\cdot)+ \left\{\prod_{h=1}^{b_0+b_1}(1-v_h)\right\}  \sum_{j=b_0+b_1+b_2+1}^{\infty}\left[ v_j \prod_{h=b_0+b_1+b_2+1}^{j-1}(1-v_h)\right]\delta_{\theta_j}(\cdot),\\
&p_2 (\cdot)= \sum_{j=1}^{b_0}\left[ v_j \prod_{h=1}^{j-1} (1-v_h)\right]\delta_{\theta_j} (\cdot)+ \left\{\prod_{h=1}^{b_0}(1-v_h)\right\}  \sum_{j=b_0+b_1+1}^{\infty}\left[ v_j \prod_{h=b_0+b_1+1}^{j-1}(1-v_h)\right]\delta_{\theta_j}(\cdot)
\end{align*}
This formulation can be extended to any number of groups $G$ by assuming that, for $j=b_0+b_1+\dots+b_{g-1}+1,\dots, b_0+b_1+\dots+b_{g}$, $\ell_{j,g}$ is equal to one for distribution $g$ and zero for the others.

The analytical tractability of the eventually single-atom structure is preserved in this symmetric version, as evidenced by the simple form of the correlation between the distributions $p_1$ and $p_2$:
\begin{equation*}
	\mathrm{Corr}(p_1(A),p_2(A); b_{0:2}) = 1-\left( \frac{\alpha}{\alpha+2} \right)^{b_0} \left[ 1-\left( \frac{\alpha}{\alpha+1} \right)^{b_1+b_2}   \right].
\end{equation*}

The correlation depends on the sequences of thinning variables solely through the number of initial shared atoms $b_0$ and the sum of $b_1$ and $b_2$, which represents the total number of atoms not shared by $p_1$ and $p_2$. 
Indeed, when $b_1 = b_2 = 0$, the correlation equals one, as $p_1 = p_2$. 

\subsubsection{Symmetric eventually single-atom Poisson thinning}\label{subsec:blocked_poisson}
One can assign independent Poisson prior distributions to all parameters modeling the length of each block, $b_r\sim \mathrm{Poisson}(\lambda_r)$ for $r=0,1,2$. Then, the marginal correlation is obtained as 
\begin{align*}
	\mathrm{Corr}(p_1,p_2)= 1- \exp\left\{-\frac{2\lambda_0}{\alpha+2}\right\} \left[ 1-  \exp\left\{-\frac{\lambda_1+\lambda_2}{\alpha+1}\right\}\right].
\end{align*}
If $\lambda_1 = \lambda_2=0$, there are no distribution-specific atoms, and the correlation is one. 

\setcounter{equation}{0}
\section{Correlations with parent process}

We consider a vector $p_{1:G}$ of thinned-DDPs obtained by thinning a common Dirichlet process $p$ (see Definition~\ref{def:thinned-DDP}), here referred to as the \say{parent process}. Focusing without loss of generality on the first component $p_1$, we study its correlation with the parent process $p$. The results below are direct specializations of those in Sections~\ref{sec::definition} and~\ref{sec:prior}, obtained by replacing $p_2$ with $p$, which corresponds to setting $\ell_2$ as an infinite sequence of ones.\\[-15pt]

A direct application of Proposition~\ref{prop:corr} leads to the following corollary.
\begin{corollary}\label{cor:corr_parent}
    If $p_{1}\sim\text{thinned-DDP}(\alpha,P_0,\ell_{1})$, with parent process $p$, then, for any $A\in\mathcal{B}$, the correlation between $p_{1}(A)$ and $p(A)$ is given by
\begin{align*}
\mathrm{Corr}\left(p_1(A), p(A) ; \ell_{1} \right) &=\frac{2}{(\alpha+2)} \sum_{j\geq 1}  \ell_{j,1}  \prod_{h<j} \frac{\alpha}{\alpha+\ell_{h,1}+1}\\
	&= \frac{2}{(\alpha+2)} \sum_{j\geq 1}  \ell_{j,1} \alpha^{j-1} \left( \frac{1}{\alpha+2} \right)^{\sum_{h=1}^{j-1} \ell_{h,1}} \left( \frac{1}{\alpha+1} \right)^{\sum_{h=1}^{j-1} (1-\ell_{h,1})}.
\end{align*}
\end{corollary}
\vspace{0.35cm}

A direct application of Corollary~\ref{corollary:corr_eventually} leads to the following corollary.
\begin{corollary}\label{cor:parent_eventually}
If $p_{1}\sim\text{thinned-DDP}(\alpha,P_0,u_{1})$ with parent process $p$, then, for any $A\in\mathcal{B}$, the correlation between $p_{1}(A)$ and $p(A)$ is given by
\begin{equation*}
	\mathrm{Corr}(p_1(A),p(A); u_{1}) = 
	\left(\frac{\alpha}{\alpha+1}\right)^{ u_1-1 }.
\end{equation*}
\end{corollary}

\vspace{0.3cm}
\noindent A direct application of Proposition~\ref{prop:corr_bernoulli} leads to the following corollary.

\begin{corollary}\label{cor:corr_parent_bernoulli}
Let $p_{1}\sim\text{thinned-DDP}(\alpha,P_0,\ell_{1})$ with parent process $p$, and consider a measurable set $A\in\mathcal{B}$. Assume $\ell_{j,1}\simiid \Bern(\pi_1)$, for $j \geq 1$. The correlation between $p_1(A)$ and $p(A)$ is
\begin{equation*}
	\mathrm{Corr}(p_1(A),p(A))  = 
	\frac{2\, \pi_1 (\alpha+1) }{\alpha(\pi_1+1)+2}.
\end{equation*}
\end{corollary}

\vspace{0.3cm}
\noindent A direct application of Proposition~\ref{prop:corr_poisson} leads to the following corollary.

\begin{corollary}\label{cor:corr_parent_eventually_poisson}
Let $p_{1}\sim\text{thinned-DDP}(\alpha,P_0,u_{1})$ with parent process $p$, and consider a measurable set $A\in\mathcal{B}$. Assume $u_1 - 1 \simind \text{Poisson}(\lambda_1)$. The correlation between $p_1(A)$ and $p(A)$ is
\begin{align*}
	\mathrm{Corr}(p_1(A),p(A))  = \exp\left\{-\frac{\lambda_1}{\alpha+1}\right\}.
\end{align*}
\end{corollary}


\setcounter{equation}{0}
\section{Blocked Gibbs sampler}\label{Asec::gibbs_derivation}

Consider the Gaussian mixture model of Section~\ref{sec::posterior_inference}.
Before detailing the algorithm, it is useful to introduce a hierarchical representation of the model based on the cluster allocation variables $z_{i,g}\in\{1,2,\dots\}$. These variables have a discrete distribution $\P(z_{i,g} = k \mid \{\omega_{j,g}\}_{j\geq1} ) = \omega_{k,g}$, for $k\geq 1$, and they are such that $f(y_{i,g}\mid z_{i,g} = k, \{\mu_j,\sigma^2_j\}_{j\geq 1}) = \phi(y_{i,g}\mid \mu_k,\sigma^2_k)$.
With this model representation, the blocked Gibbs sampler proceeds by iteratively updating each parameter from its full conditional distribution, as follows
\begin{enumerate}
  \item Update the thinning variables $\{\ell_{j,g}\}_{j=1}^{T}$, $g=1,\dots,G$.\\
  Assume that the last occupied cluster is $K = \max\{j: n_{j,g}>0,\, g=1,\dots,G\}$, where $n_{j,g}$ is the number of observations of group $g$ allocated to cluster $j$, for $g=1,\dots,G$ and $j\geq 1$. 
  For $k = K, K-1,\dots, 1$, and for $g= 1,\dots,G$, update the thinning variable $\ell_{k,g}$:
  \begin{equation*}
    \ell_{k,g}\mid \dots \sim \begin{cases}
    \delta_{1} \quad \text{if $n_{k,g} > 0$}\\
    \text{Bernoulli}(\pi^\star) \quad \text{if $n_{k,g} = 0$}
    \end{cases}
  \end{equation*}
  with 
  \begin{equation*}
    \pi^\star = D^{-1} \big(1-v_{k}\big)^{\sum_{h=k+1}^K n_{h,g}} \pi_g
  \end{equation*}
  and $D = \big(1-v_{k}\big)^{\sum_{h=k+1}^K n_{h,g}} \pi_g + 1-\pi_g$.	
  
  Sample a new thinning variable from the prior, $\ell_{k,g}\sim \text{Bernoulli}(\pi_g)$, for $k = K+1, K+2, \dots, T$, where $T$ is the truncation level.
  \item Update the beta random variables $\{v_j\}_{j=1}^{\infty}$.
  For $k=1,2,\dots,T$
  \begin{equation*}
 v_{k}\mid \dots \sim \Beta \left( 1+ \sum_{g=1}^G n_{k,g} ,\:\: \alpha + \sum_{g=1}^G \bigg\{ \ell_{k,g} \bigg(\sum_{h=k+1}^K n_{h,g} \bigg)\bigg\} \right).
  \end{equation*}
  With the updated variables, compute the weights via a thinned-SB process.
  \item Update the cluster allocation ${z_{i,g}}$.\\	
  For $g=1,\dots,G$ and $i=1,\dots,N_g$, sample ${z_{i,g}}$ from a multinomial distribution 
	\begin{equation*}
		\P(z_{i,g}=k \mid \dots ) \propto \begin{cases}
			\phi(y_{i,g}\mid\mu_k, \sigma^2_k)  \bigg[v_k \prod_{h=1}^{k-1}(1-\ell_{h,g}v_{h})\bigg] \quad \text{if } \ell_{k,g} = 1 \\
			0\quad \text{if } \ell_{k,g} = 0
		\end{cases}
	\end{equation*}
 for $k=1,\dots,T$.
  \item Sample the kernel parameters $\{\mu_k, \sigma^2_k\}_{k=1}^{\infty}$.\\
  For $k=1,2,\dots,T$, denote with $A_k$ the set $ A_k = \{ (g,i) : z_{i,g} = k \}$ and with $n_k$ its cardinality, $n_{k} = \sum_{g=1}^{G}\sum_{i=1}^{N_g} \mathbb{I}( z_{i,g} = k )$, sample 
  \begin{equation*}
	(1/\sigma^2_k \mid \dots)\sim \textrm{Gamma} \left( \gamma_0 + \frac{n_k}{2}, \lambda_0 + \frac{1}{2}\bigg\{ \sum_{A_k}(y_{i,g} - \bar{y}_{A_k})^2 + \frac{\tau_0 n_k}{\tau_0 + n_k}(\bar{y}_{A_k} - \mu_0)^2 \bigg\} \right)
  \end{equation*}
  \begin{equation*}
	(\mu_k \mid \sigma^2_k,\dots)\sim \N \left(  \frac{\tau_0\mu_0 + \sum_{A_k}y_{i,g}}{\tau_0 + n_k}, \frac{\sigma_k^2}{\tau_0+n_k} \right)
  \end{equation*}
  where $\bar{y}_{A_k} = n_k^{-1}\sum_{A_k}y_{i,g}$.
  \item Update the distribution-specific thinning probabilities $\pi_g$.\\
  For $g=1,\dots,G$, sample a value from a beta distribution with updated parameters,
  \begin{equation*}
  	\pi_g\mid \dots \sim \text{Beta}\left(a_{\pi}+\sum_{j=1}^{T} \ell_{j,g}, b_{\pi} + T - \sum_{j=1}^{T} \ell_{j,g}\right).
  \end{equation*}
\end{enumerate}
\setcounter{equation}{0}
\section{Additional details on the simulation studies}\label{Asec::simu}

Estimation of the thinned-DDP was performed using the Gibbs sampler outlined in Section~\ref{sec::posterior_inference}; for the complete-pooling and no-pooling models, we used a standard blocked Gibbs sampler for DP mixtures. The truncation parameter was set to 50 for all models. 
The concentration parameter $\alpha$ for the thinned-DDP, the complete-pooling and no-pooling mixtures, and the GM-DDP was fixed to 1, while it is random for the CAM. The thinning probabilities $\pi_g$, for $g=1,\dots,G$, in the thinned-DDP were random and assigned a $\text{Beta}(3, 3)$ prior distribution. 
The parameters of the normal-inverse gamma base measure were fixed to $(\mu_0 = \bar{y}, \tau_0 = 0.01, \gamma_0 = 2.5, \lambda_0 = 1.5)$. 
Convergence was assessed by analyzing the effective sample size of the number of occupied clusters and the deviance of the estimated density. The latter is defined as $$ D = -2 \sum_{g=1}^G \sum_{i=1}^{n_g} \log \sum_{j} \frac{m_{j,g}}{n_g} \phi(y_{i,g}\mid \mu_j,\sigma^2_j),$$
where $m_{j,g}$ is the number of observations in group $g$ allocated to cluster $j$. 
These summary quantities are related to the inferential goals and are invariant to label permutations \citep{canale2022importance, kalli2011slice, papaspiliopoulos2008retrospective, neal2000markov}.

\begin{figure}[h!]
	\centering
	\includegraphics[width=.78\linewidth]{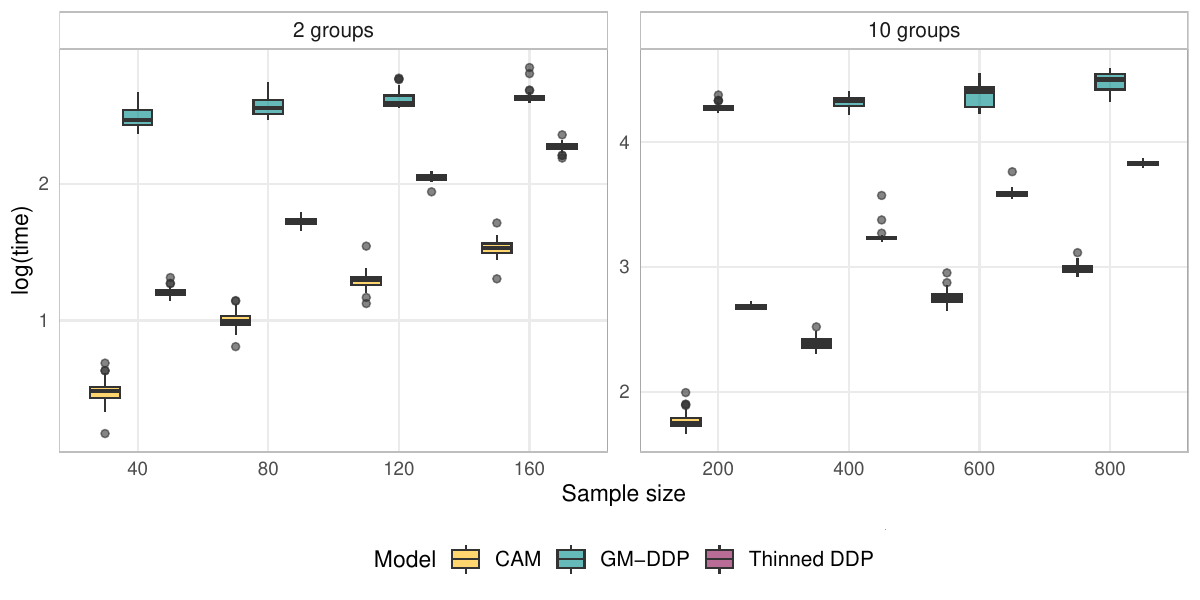}
	\caption{Distribution of the computational time for the thinned-DDP, the CAM, and the GM-DDP mixture on the simulated data. }\label{fig::time}
\end{figure}

Posterior inference for the CAM and the GM-DDP was performed using the R packages sanba \citep{sanba} and BNPmix \citep{BNPmix}, respectively.
We ran all algorithms for 10,000 iterations, discarding the first 5,000 as burn-in, and thinned the chain by a factor of 2. Figure~\ref{fig::time} shows the distribution of the computation time (in logarithmic scale) for the thinned-DDP, the CAM, and the GM-DDP. We do not report the computational time of the HDP algorithm because its implementation is in R, whereas all competing methods are implemented in C++. As a result, its execution time is orders of magnitude larger and not directly comparable. 

\subsection{Sensitivity analysis}
We conducted a sensitivity analysis to evaluate the effect of the prior specification for the thinning probabilities on posterior inference. Throughout the article, results are reported under the prior specification $\pi_g \overset{iid}{\sim} \mathrm{Beta}(3,3)$, $g=1,\ldots,G$, which induces a mildly informative symmetric prior centered at 0.5.
To assess the robustness of our findings, we also considered three alternative prior distributions: a $\mathrm{Beta}(10,10)$, which remains centered at 0.5 but is substantially more concentrated; a $\mathrm{Beta}(1,1)$, corresponding to the uniform distribution on $(0,1)$; and a $\mathrm{Beta}(0.1,0.1)$, which places most of its mass near the boundaries of the parameter space, thereby favoring values close to 0 and 1.

\begin{figure}[hb]
	\centering
	\includegraphics[width=.78\linewidth]{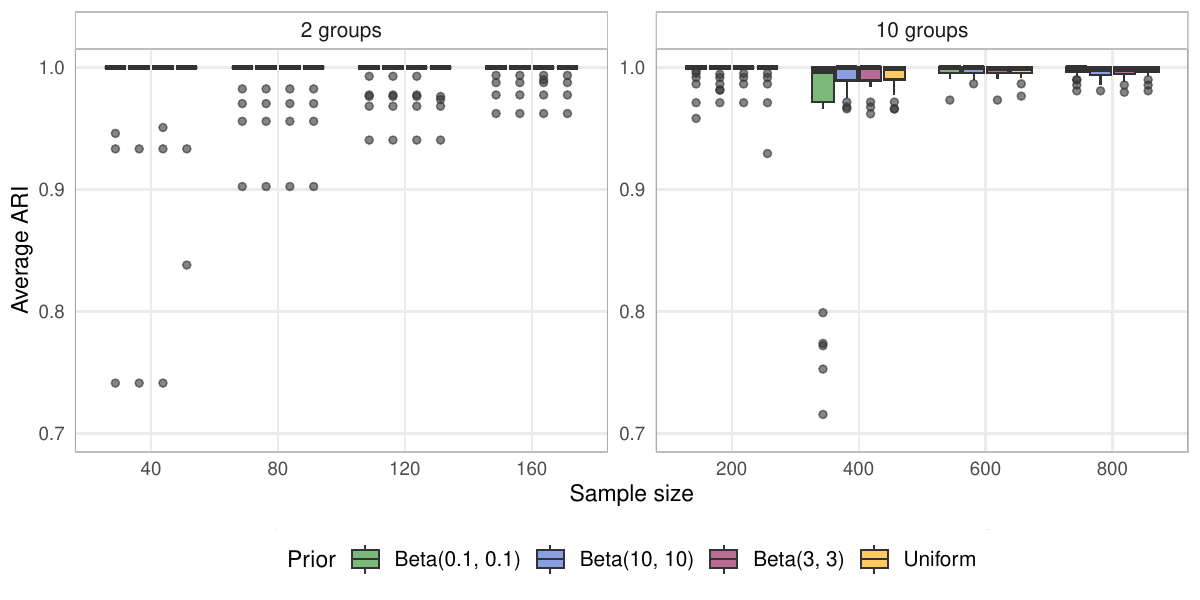}
	\caption{Boxplot of the average ARI for the thinned-DDP mixture under the different prior specifications on the simulated data.  \label{fig::ARI_sensit}}
\end{figure}

\begin{figure}[t]
	\centering
	\includegraphics[width=.78\linewidth]{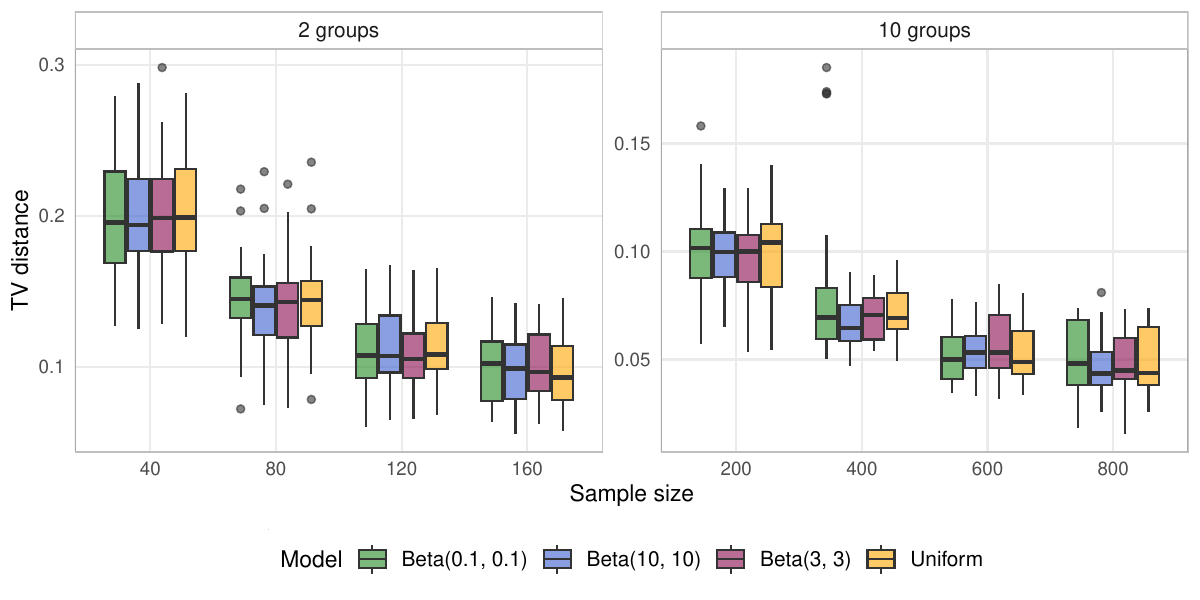}
	\caption{Boxplot of the TV distance for the thinned-DDP mixture under the different prior specifications on the simulated data. \label{fig::TV_sensit}}
    \vspace{2cm}
\end{figure}

We considered the synthetic data of Section~\ref{sec::simulation}, and fitted the thinned-DDP with these prior specifications on 25 replicates. Figures~\ref{fig::ARI_sensit} and~\ref{fig::TV_sensit} show the distributions of the average ARI and TV distance for the four prior specifications. The results are very stable across scenarios, demonstrating the robustness of the model to different prior choices. 

\vspace{1cm}

\clearpage

\subsection{Additional figures} 
\begin{figure}[bh!]
	\centering
	\includegraphics[width=.88\linewidth]{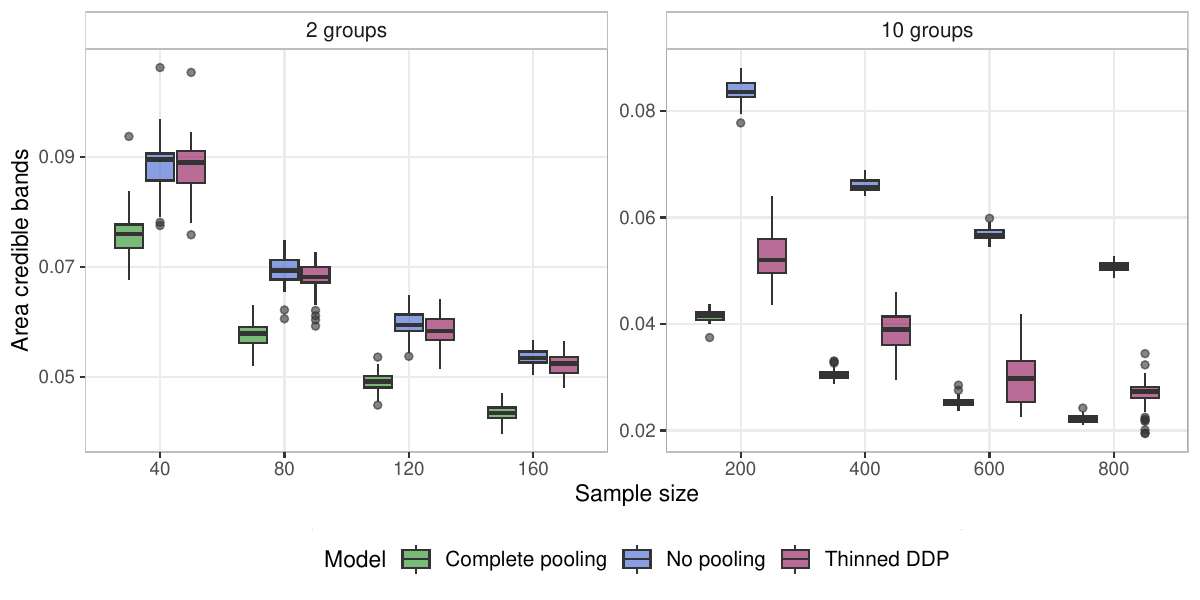}
	\caption{Distribution of the average pointwise size of the credible bands for the thinned-DDP, the complete-pooling, and the no-pooling DP mixture on the simulated data. \label{fig::areaCI1}}
\vspace{1.5cm}
	\centering
	\includegraphics[width=.88\linewidth]{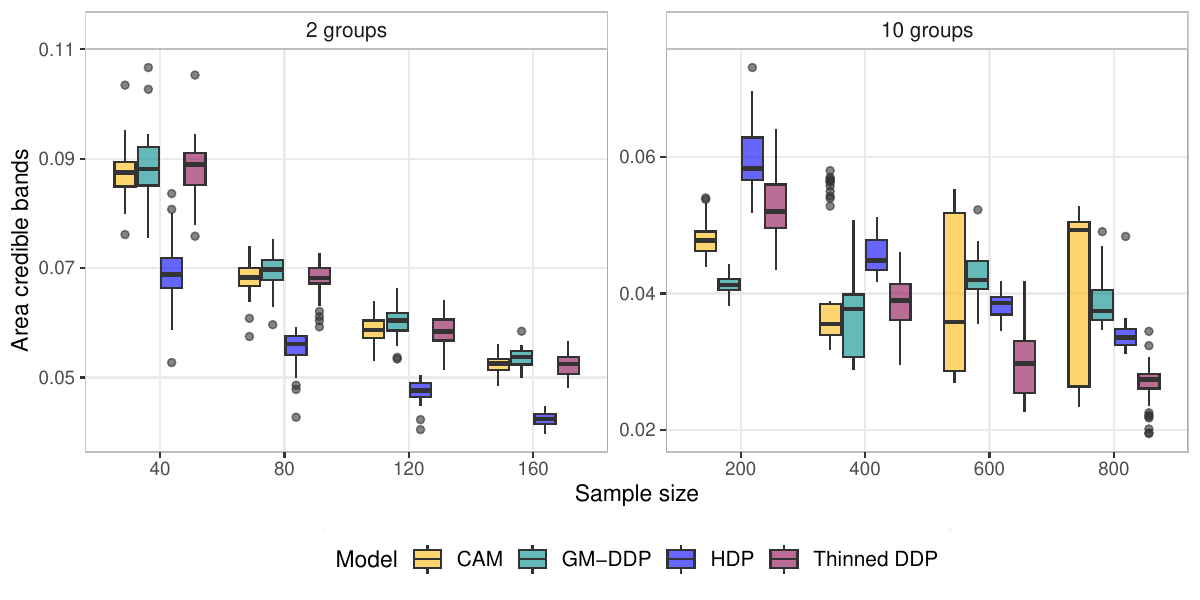}
	\caption{Distribution of the average pointwise size of the credible bands for the thinned-DDP, the CAM, the GM-DDP, and the HDP mixture on the simulated data. \label{fig::areaCI2}}
\end{figure}

Figure~\ref{fig::areaCI1} displays a summary of the distribution of the average length of the pointwise 95\% HPD credible intervals for the thinned-DDP, the complete pooling, and no pooling mixtures. Figure~\ref{fig::areaCI2} displays the same quantity for the thinned-DDP, the GM-DDP, and the CAM.

\begin{figure}[ht!]
	\centering
	\includegraphics[width=0.9\linewidth]{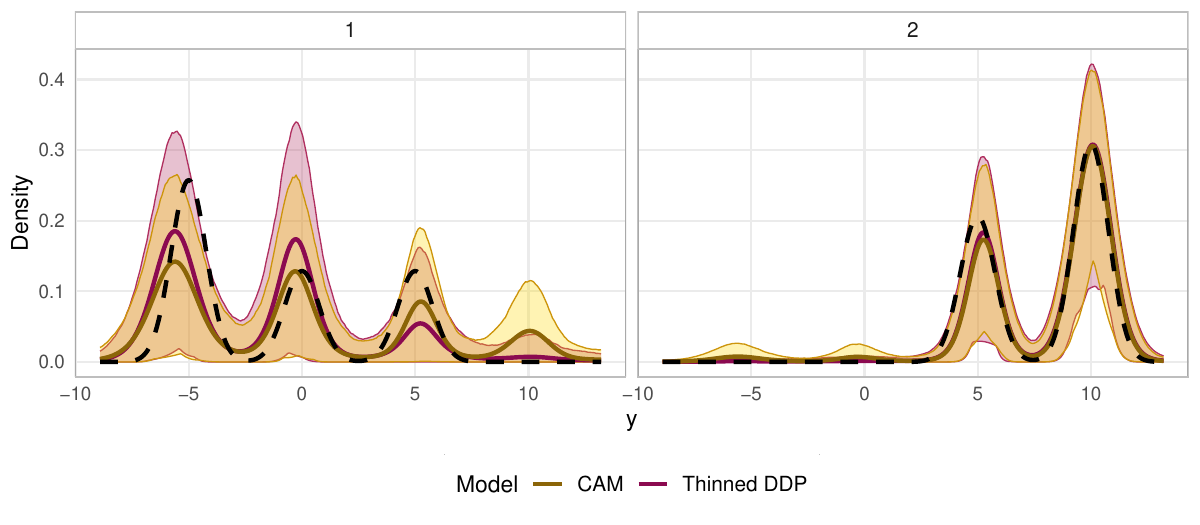}
    \vspace{.5cm}
	\centering
	\includegraphics[width=0.9\linewidth]{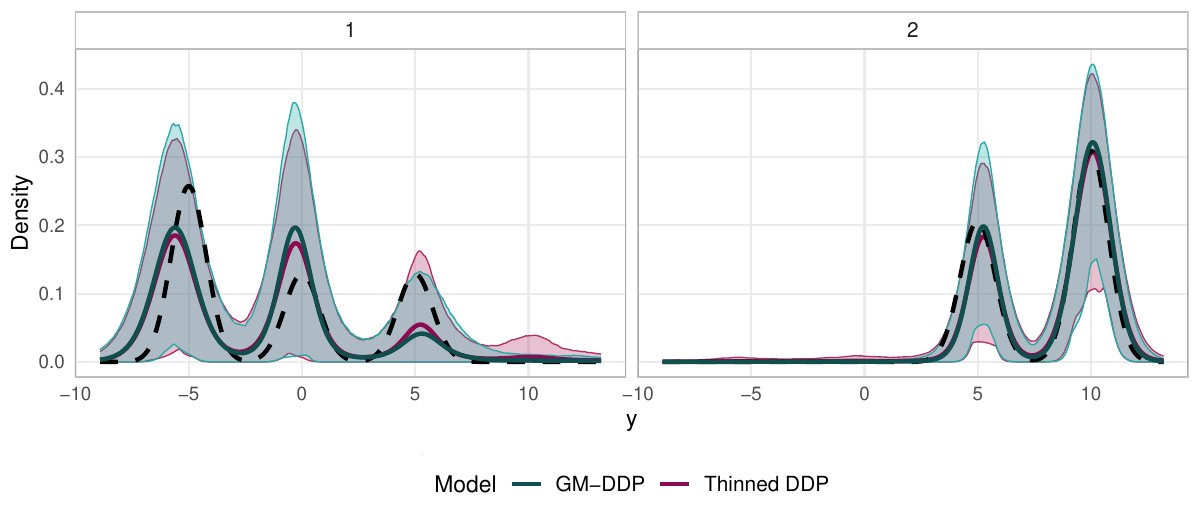}
    \vspace{.5cm}
	\centering
	\includegraphics[width=0.9\linewidth]{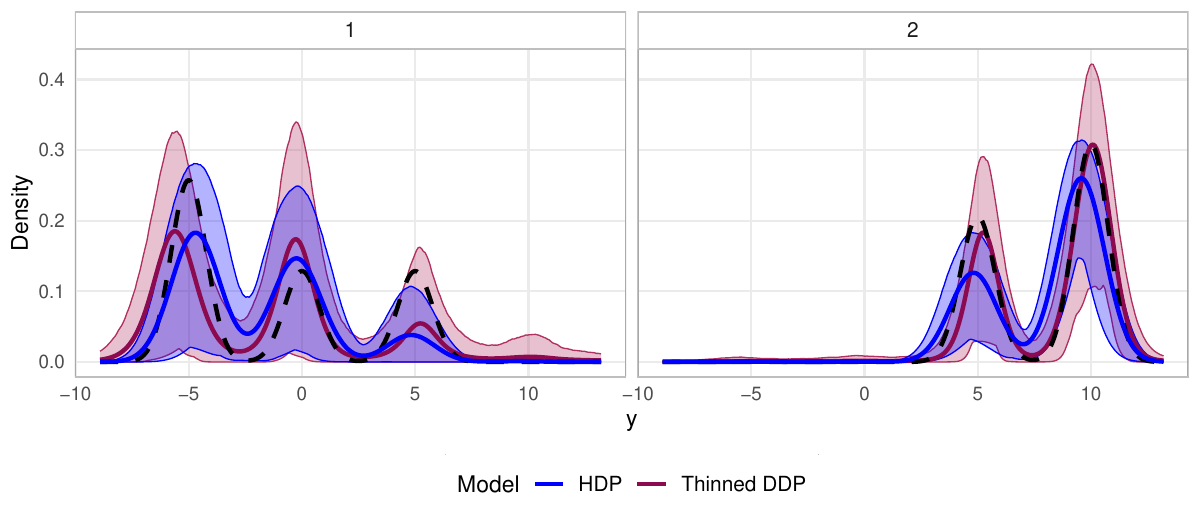}
	\caption{$G=2$, $N=40$: posterior density estimates (continuous lines), and credible bands (shaded areas). The black dashed line represents the true data-generating density. Top: thinned-DDP and CAM; center: thinned-DDP and GM-DDP; bottom: thinned-DDP and HDP. }\label{fig::density_est_2G_minSS}
\end{figure}

\begin{figure}[h!]
	\centering
	\includegraphics[width=.9\linewidth]{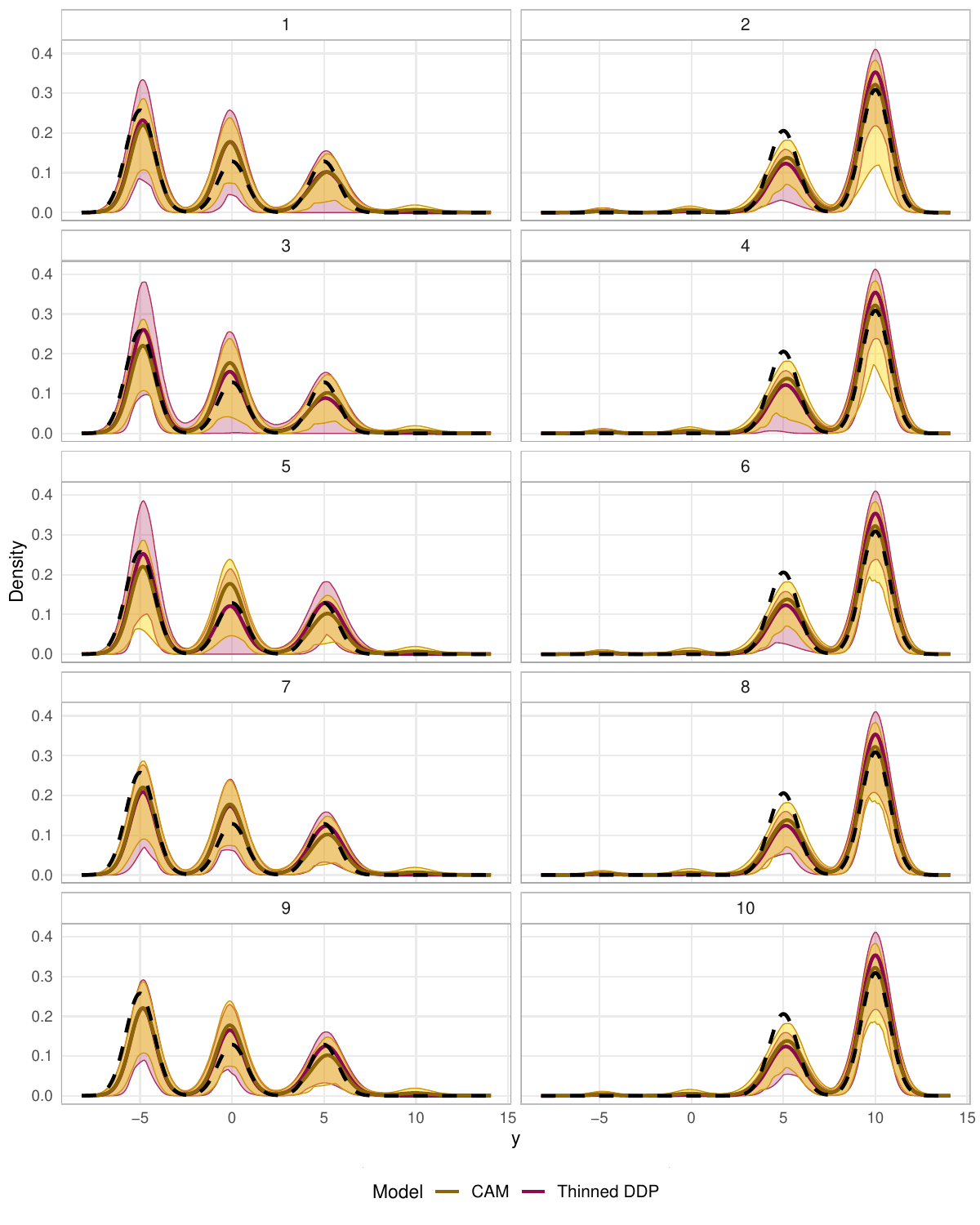}
	\caption{$G=10$, $N=200$: posterior density estimates (continuous lines), and credible bands (shaded areas). The black dashed line represents the true data-generating density. Estimates of the thinned-DDP and the CAM}\label{fig::density_est_10G_minSS_cam}
\end{figure}

\begin{figure}[h!]
	\centering
	\includegraphics[width=.9\linewidth]{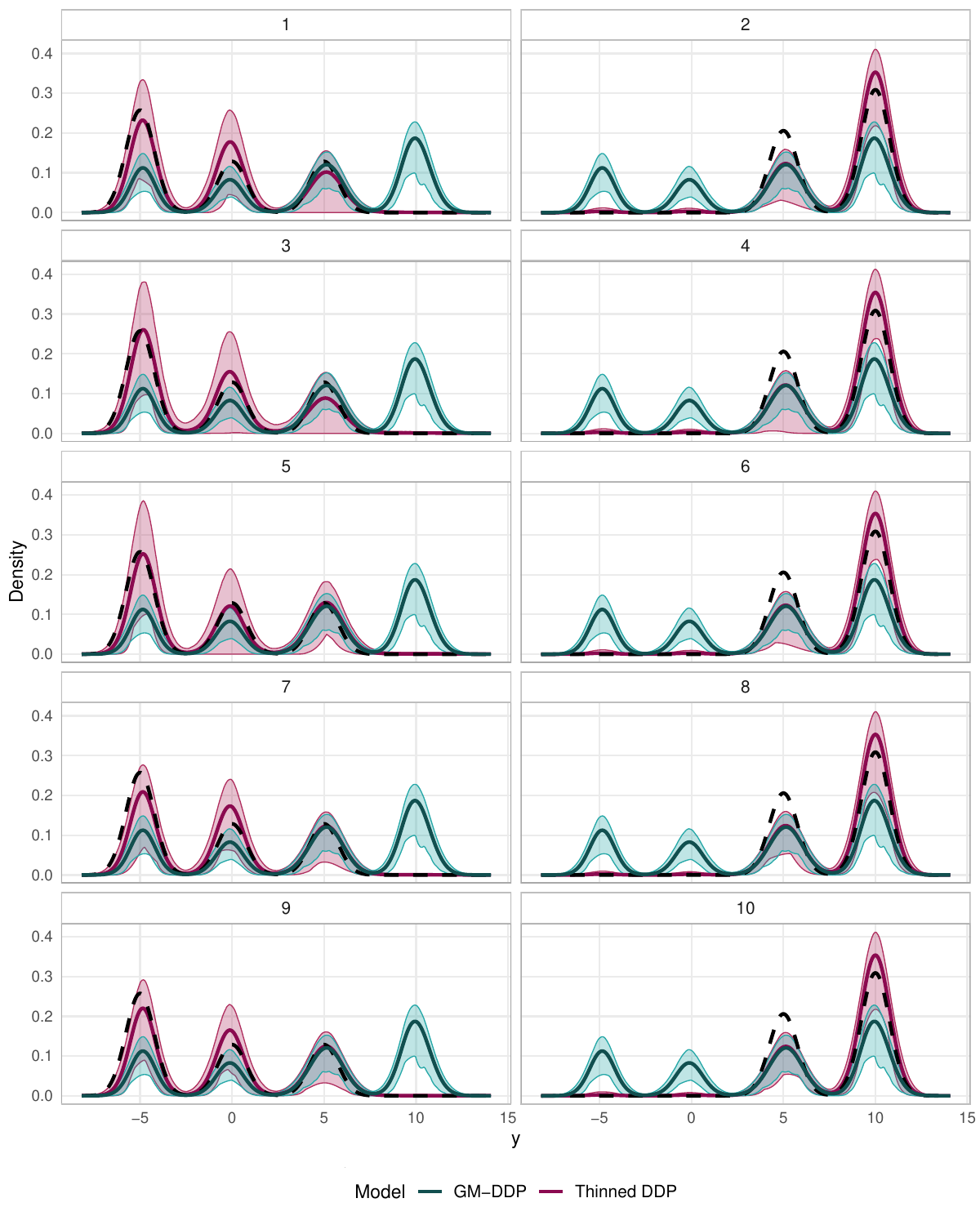}
	\caption{$G=10$, $N=200$: posterior density estimates (continuous lines), and credible bands (shaded areas). The black dashed line represents the true data-generating density. Estimates of the thinned-DDP and the GM-DDP.} \label{fig::density_est_10G_minSS_gm}
\end{figure}

\begin{figure}[h!]
	\centering
	\includegraphics[width=.9\linewidth]{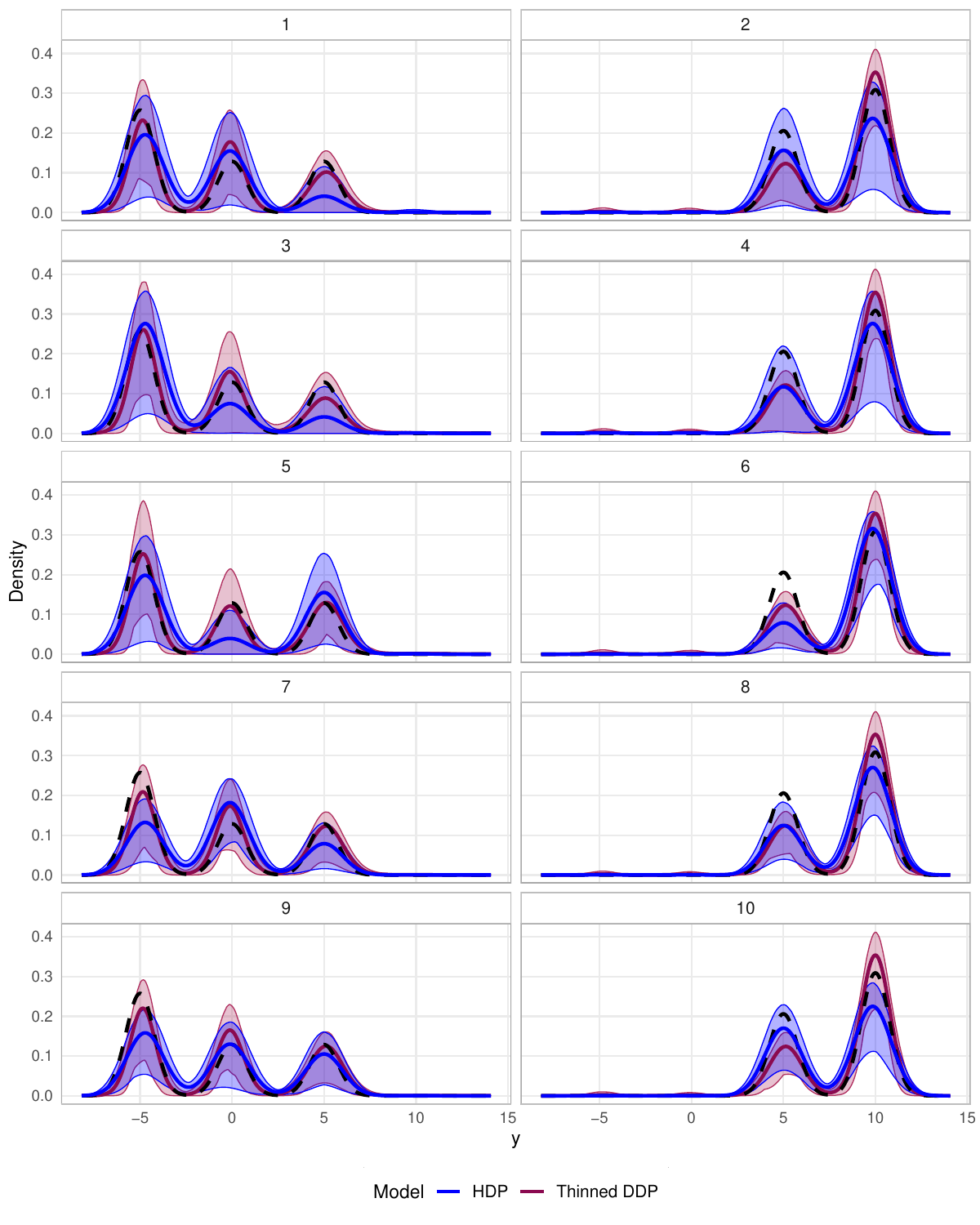}
	\caption{$G=10$, $N=200$: posterior density estimates (continuous lines), and credible bands (shaded areas). The black dashed line represents the true data-generating density. Estimates of the thinned-DDP and the HDP.} \label{fig::density_est_10G_minSS_hdp}
\end{figure}

\begin{figure}[ht!]
	\centering
	\includegraphics[width=0.9\linewidth]{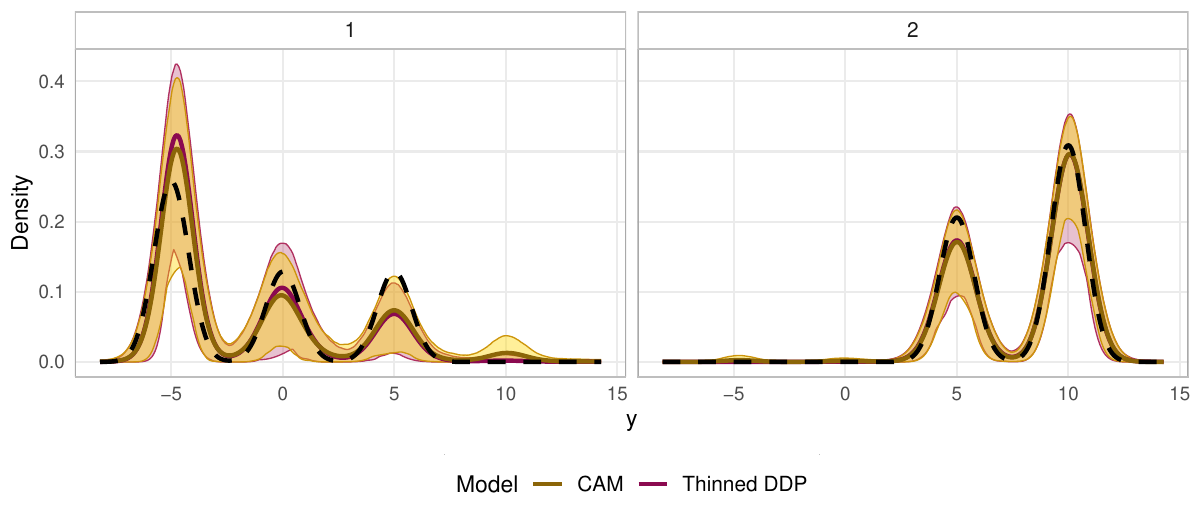}
    \vspace{.1cm}
	\centering
	\includegraphics[width=0.9\linewidth]{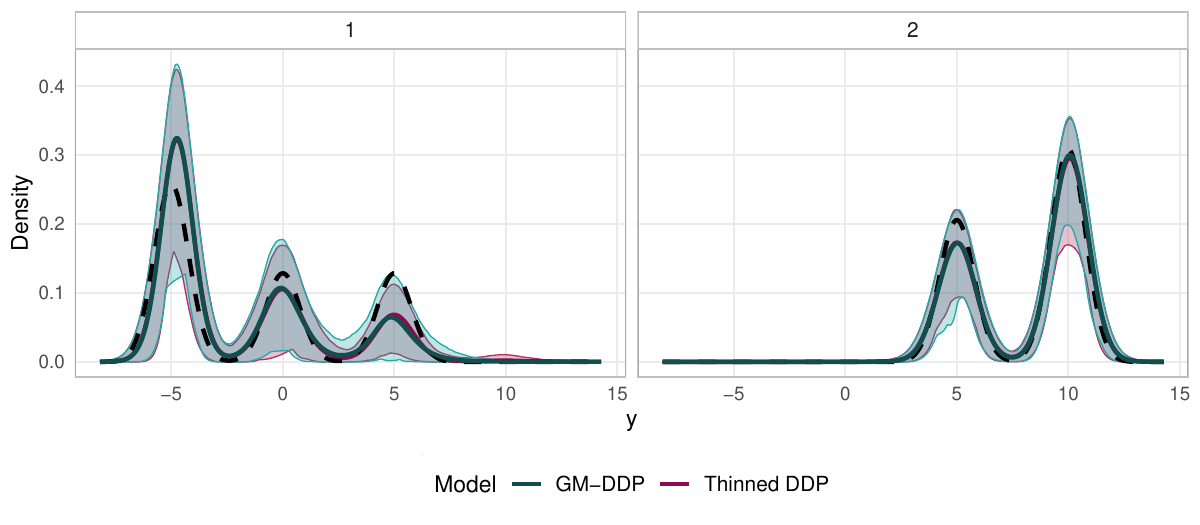}
    \vspace{.1cm}
	\centering
	\includegraphics[width=0.9\linewidth]{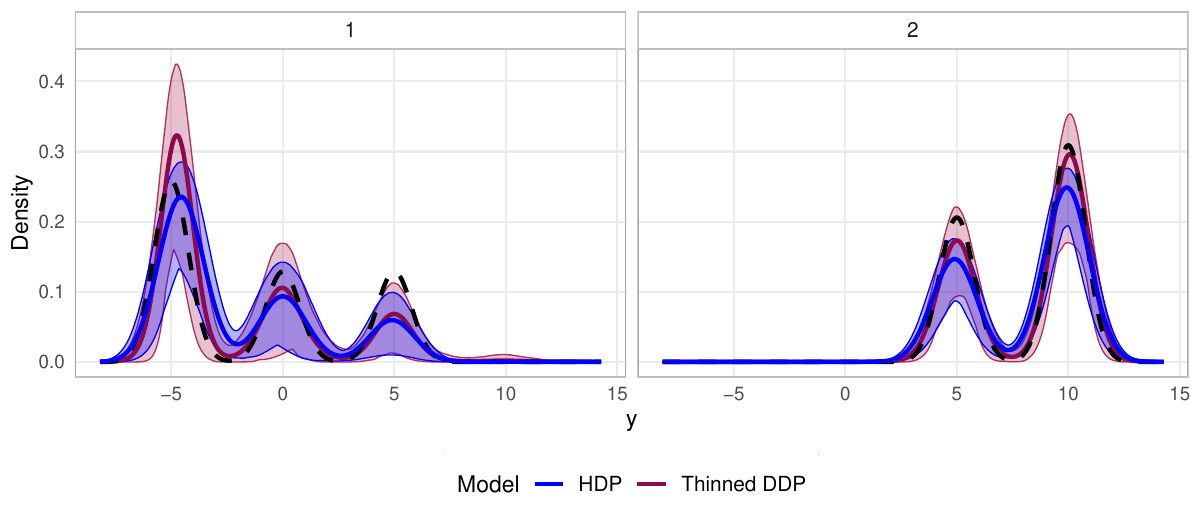}
	\caption{$G=2$, $N=160$: posterior density estimates (continuous lines), and credible bands (shaded areas). The black dashed line represents the true data-generating density. Top: thinned-DDP and CAM; center: thinned-DDP and GM-DDP; bottom: thinned-DDP and HDP. }\label{fig::density_est_2G_maxSS}
\end{figure}

\begin{figure}[h!]
	\centering
	\includegraphics[width=.9\linewidth]{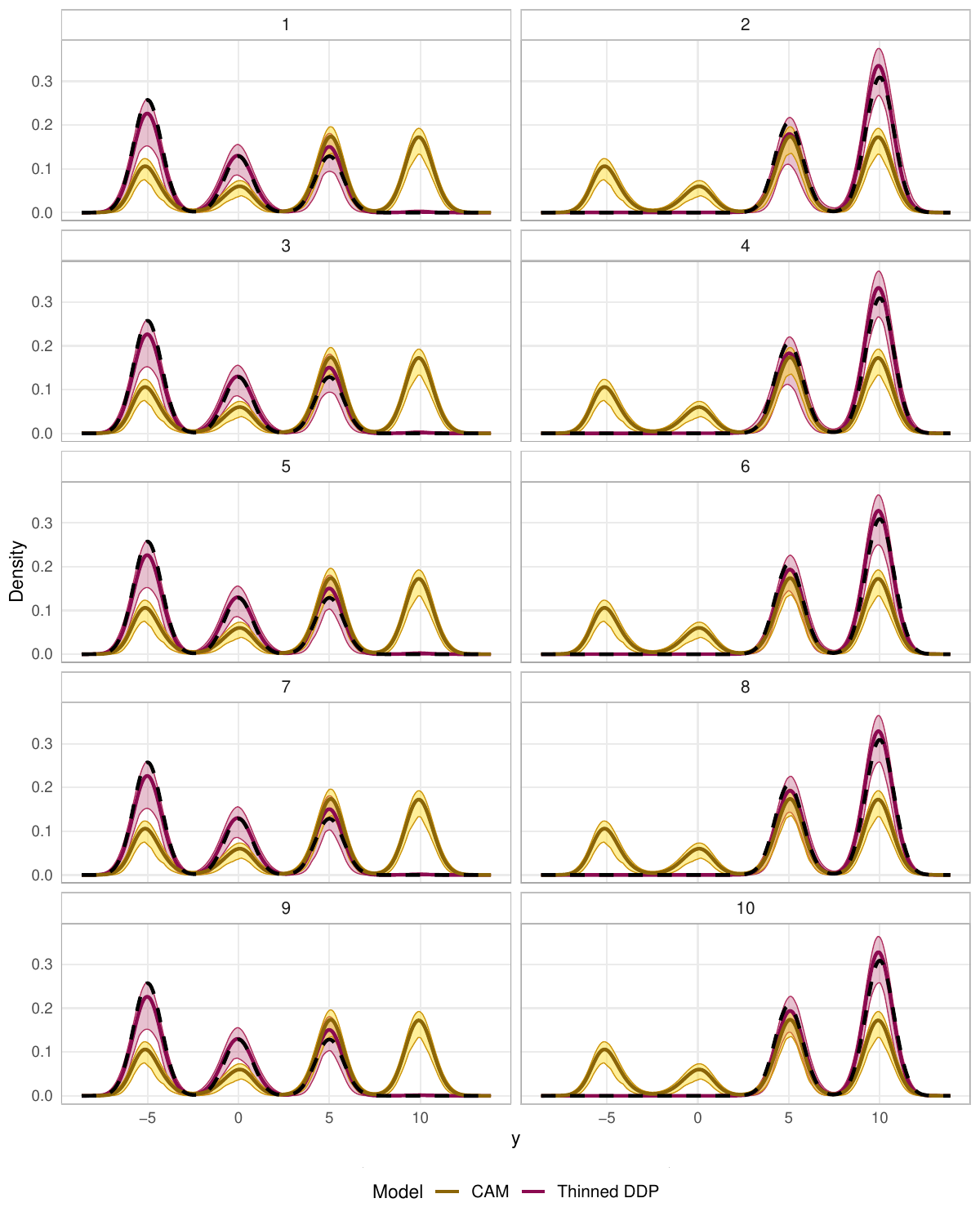}
	\caption{$G=10$, $N=800$: posterior density estimates (continuous lines), and credible bands (shaded areas). The black dashed line represents the true data-generating density. Estimates of the thinned-DDP and the CAM}\label{fig::density_est_10G_maxSS_cam}
\end{figure}

\begin{figure}[h!]
	\centering
	\includegraphics[width=.9\linewidth]{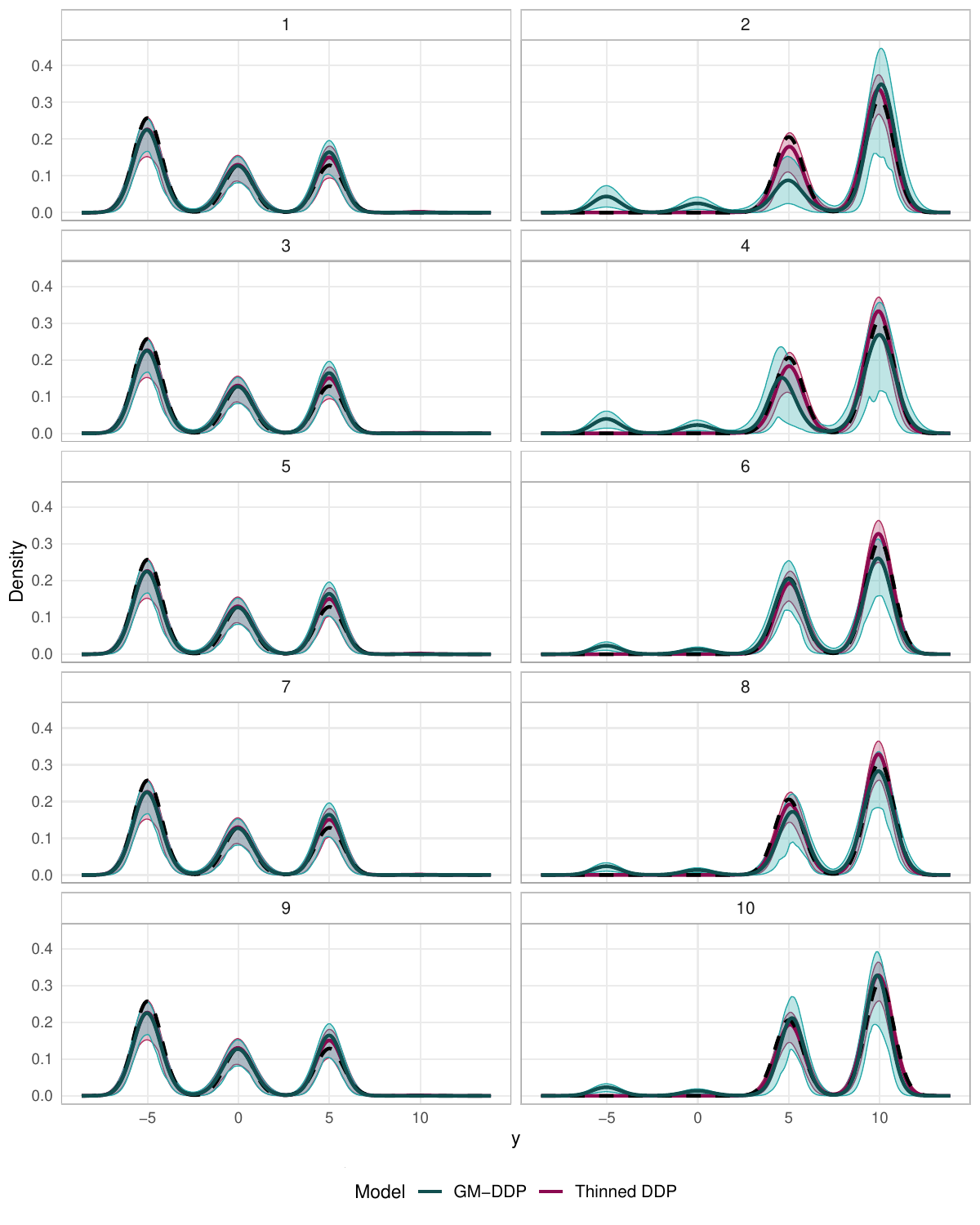}
	\caption{$G=10$, $N=800$: posterior density estimates (continuous lines), and credible bands (shaded areas). The black dashed line represents the true data-generating density. Estimates of the thinned-DDP and the GM-DDP.} \label{fig::density_est_10G_maxSS_gm}
\end{figure}

\begin{figure}[h!]
	\centering
	\includegraphics[width=.9\linewidth]{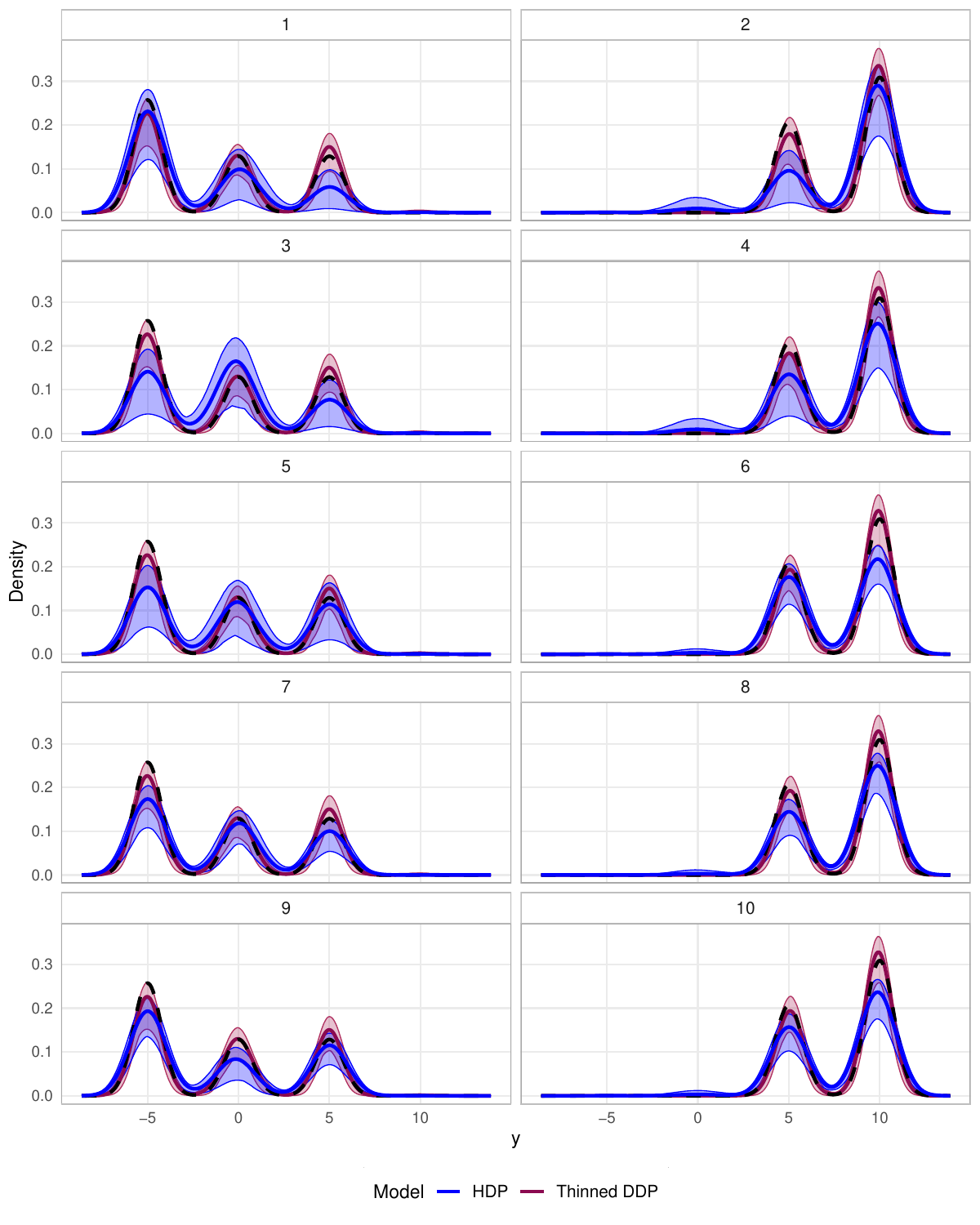}
	\caption{$G=10$, $N=800$: posterior density estimates (continuous lines), and credible bands (shaded areas). The black dashed line represents the true data-generating density. Estimates of the thinned-DDP and the HDP.} \label{fig::density_est_10G_maxSS_hdp}
\end{figure}

Finally, we report additional graphs with the posterior point estimates of the densities and the credible bands for the thinned-DDP, the CAM, the GM-DDP, and the HDP. Specifically, we show an example with $G=2$ and $G=10$ and the smallest sample size (Figures~\ref{fig::density_est_2G_minSS},~\ref{fig::density_est_10G_minSS_cam},~\ref{fig::density_est_10G_minSS_gm} and~\ref{fig::density_est_10G_minSS_hdp}); and with $G=2$ and $G=10$, for the largest sample size (Figures~\ref{fig::density_est_2G_maxSS},~\ref{fig::density_est_10G_maxSS_cam},~\ref{fig::density_est_10G_maxSS_gm} and~\ref{fig::density_est_10G_maxSS_hdp}).

\clearpage
\setcounter{equation}{0}
\section{Additional details on the real data analysis}\label{Asec::appl}
\begin{figure}[h!]
	\centering
	\includegraphics[width=\linewidth]{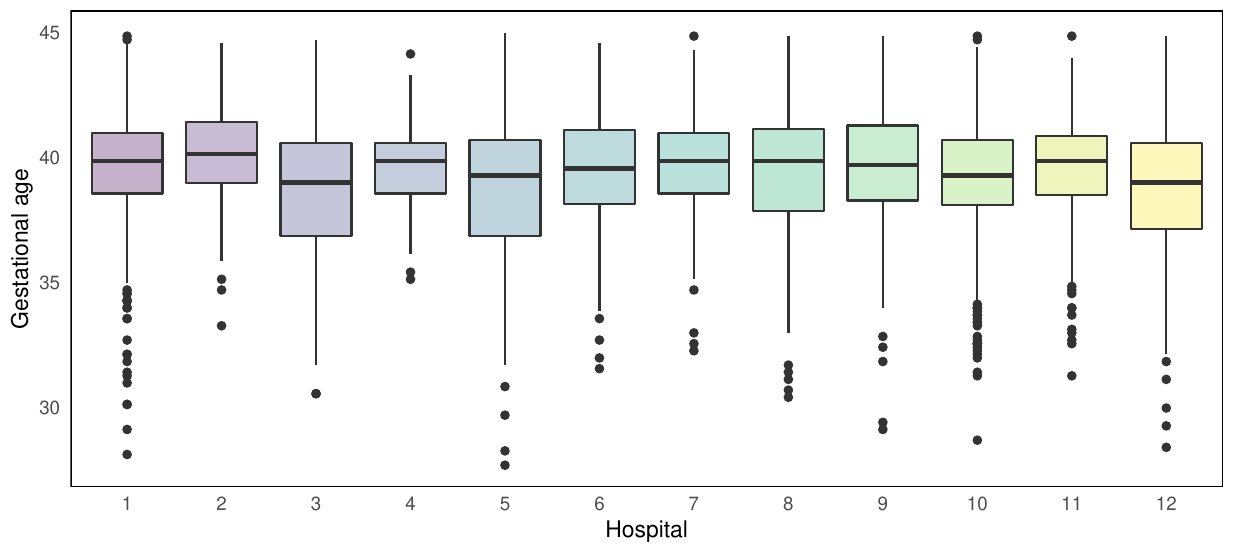}
	\caption{Distribution of the gestational age in	 the 12 centers.}\label{fig::boxplots}
\end{figure}

Figure~\ref{fig::boxplots} shows the boxplots of the distribution of the gestational age in the 12 hospitals.

\begin{figure}[h]
	\centering
	\includegraphics[width=.4\linewidth]{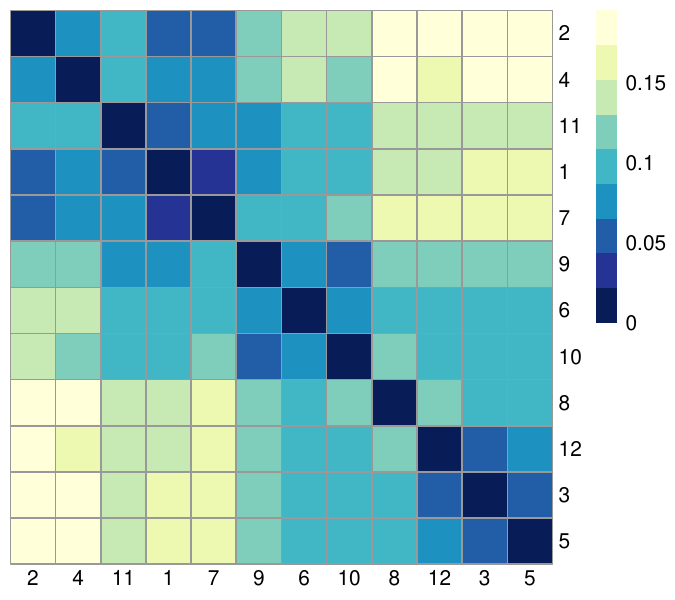}
	\caption{Heatmap of the pairwise TV distance between couples of hospitals.}\label{fig::TV_CPP}
\end{figure}

Figure~\ref{fig::TV_CPP} shows the estimated total variation distance computed on the group-specific densities between all pairs of hospitals.

\end{document}